\documentstyle[aps,pre,eqsecnum,epsf]{revtex}
 
\begin{document}
  \def\OP{\raisebox{.2ex}{$\stackrel{\leftrightarrow}{\bf P}$}}
  \renewcommand{\thesection}{\arabic{section}}
 \twocolumn[{
  \title{{\rm PHYSICAL REVIEW E}\hfill {\sl Submitted}\\~~\\
  Exact  Resummations in the Theory of Hydrodynamic Turbulence:\\
  III. Scenarios for Anomalous Scaling and Intermittency}
  \author {Victor L'vov\cite{lvov}  and Itamar  Procaccia\cite{procaccia}  }
  \address{Departments of~~\cite{lvov}Physics of Complex Systems
  {\rm and}~~\cite{procaccia}Chemical Physics,\\
   The Weizmann Institute of Science,
  Rehovot 76100, Israel,\\
  \cite{lvov}Institute of Automation and Electrometry,
   Ac. Sci. of Russia, 630090, Novosibirsk, Russia
   }
 \maketitle
\widetext
 \begin{abstract}
 \leftskip 54.8pt
 \rightskip 54.8pt
 Elements of the  analytic structure of anomalous scaling and intermittency
 in fully developed hydrodynamic turbulence are described. We focus here on
 the structure functions of velocity differences that satisfy inertial
 range scaling laws $S_n(R)\sim R^{\zeta_n}$, and the correlation of energy
 dissipation $K_{\epsilon\epsilon}(R) \sim R^{-\mu}$.  The goal is to
 understand the exponents $\zeta_n$ and $\mu$ from first principles. In
 paper II of this series it was shown that the existence of an ultraviolet
 scale (the dissipation scale $\eta$) is associated with a spectrum of
 anomalous exponents that characterize the ultraviolet divergences of
 correlations of gradient fields. The leading scaling exponent in this
 family was denoted $\Delta$. The exact resummation of ladder diagrams
 resulted in the calculation of $\Delta$ which satisfies the scaling
 relation $\Delta=2-\zeta_2$. In this paper we continue our analysis and
 show that nonperturbative effects may introduce multiscaling (i.e.
 $\zeta_n$ not linear in $n$) with the renormalization scale being
 the infrared outer scale of turbulence $L$.  It is shown that deviations
 from K41 scaling of $S_n(R)$ ($\zeta_n\neq n/3$)  must appear if the
 correlation of dissipation is mixing (i.e. $\mu>0$).  We derive an exact
 scaling relation  $\mu = 2\zeta_2-\zeta_4$. We present  analytic
 expressions  for $\zeta_n$ for all $n$ and discuss their relation to
 experimental data. One surprising prediction is that the time decay
 constant $\tau_n(R)\propto R^{z_n}$ of $S_n(R)$ scales independently of
 $n$:  the dynamic scaling exponent $z_n$ is the same for all $n$-order
 quantities, $z_n=\zeta_2$.
 \end{abstract}
 \leftskip 54.8pt
}]
\narrowtext
 \section{Introduction}
 This paper is the last in a series of four
 papers\cite{95LP-a,95LP-b,95LP-c}.  The aim of these four papers is to lay
 out the analytic basis for the description of scaling properties in fully
 developed hydrodynamic turbulence. We are concerned here with the
 statistical properties of turbulence in terms of averages over the fields
 of the fluid.  The fundamental field in hydrodynamics is the fluid's Eulerian
 velocity, denoted as ${\bf u}({\bf r},t)$ where $\bf r$ is a
 point in $d$-dimensional space (usually $d=2$ or 3) and $t$ is the time.
 The statistical quantities that have attracted decades of experimental and
 theoretical attention (see for example \cite{41Kol,MY-2,95Nel,Fri,93SK})
 are the structure functions of velocity differences, denoted as $S_n(R)$
 \begin{equation}
 S_n(R) = \left\langle\vert\hbox{\bf u}({\bf
 r}+{\bf R},t) -{\bf u}({\bf r},t)\vert^n\right\rangle \
 \label{a5}
 \end{equation}
 where $\left<\dots\right>$ stands for a suitably defined ensemble average.
 It has been asserted for a long time that the structure functions depend
 on $R$ as power laws:
 \begin{equation}
 S_n(R) \propto R^{\zeta_n} \,,
 \label{scalelaw}
 \end{equation}
 when $R$ is inside the so called ``inertial range", i.e.  $\eta
 \ll R \ll L$. Here  $\eta$  and $L$ are respectively the inner (viscous) and
outer
 (energy containing) scale of turbulence. One of the major
 questions in fundamental turbulence research is whether the scaling
 exponents $\zeta_n$ are correctly predicted by the classical Kolmogorov 41
 theory \cite{41Kol} (known as K41) in which $\zeta_n = n/3$, or whether
 these exponents deviate from $n/3$ as has been indicated by experiments.
 In particular we want to know whether the exponents $\zeta_n$ may manifest
 the phenomenon of ``multiscaling" with $\zeta_n$ being a nonlinear
 function of $n$.

 Experimental research has not confined itself to the measurement of the
 structure functions of velocity differences. Gradient fields have
 featured  as well.  For example the correlation function of the energy
 dissipation field has been studied extensively. The dissipation field
 $\epsilon({\bf r},t)$ is defined as
  \begin{equation}
 \epsilon({\bf r},t) \equiv {\nu\over 2}[\nabla_\alpha u_\beta({\bf r},t)+
 \nabla_\beta u_\alpha({\bf r},t)]^2    \label{defeps}
 \end{equation}
 with $\nu$  being the kinematic viscosity. The correlation function of the
 dissipation field $K_{\epsilon\epsilon}(R)$ is
 \begin{equation}
 K_{\epsilon\epsilon}(R) = \left< \hat\epsilon ({\bf r}+{\bf R},t)
 \hat\epsilon ({\bf r},t) \right> \ , \label{Kee}
 \end{equation}
 where  $\hat\epsilon({\bf r},t) = \epsilon ({\bf r},t) - \bar\epsilon$.
 Here and below $\bar\epsilon\equiv \left<\epsilon\right>$.  Experiments
 appeared to show \cite{MY-2,95Nel,Fri,93SK} that
$K_{\epsilon\epsilon}(R)$ decays according to a power law,
  \begin{equation}
 K_{\epsilon\epsilon}(R) \sim  R^{-\mu} \,,\quad 	\eta \ll  R \ll L
\label{Rmiu}
 \end{equation}
 with $\mu$ having a numerical value of $0.2-0.3$. The analytic derivation
 of this law from the equations of fluid mechanics and the calculation of
 the numerical value of the scaling exponent $\mu$  have been among the
 elusive goals of theoretical research.

 In paper 0 of this series (subtitled ``Line-Resummed Diagrammatic
 Perturbation Theory") we reviewed the literature (\cite{61Wyl,73MSR,78DP}
 with the aim of introducing the student to the available techniques. That
 paper did not present any new results.  In paper I of this series
 \cite{95LP-b} (subtitled ``The Ball of Locality and Normal Scaling") we
 dealt with the perturbative theory of the correlation, response and
 structure functions of hydrodynamic turbulence.  The main result of that
 paper (and see also \cite{87BL}) was that after appropriate resummations
 and renormalizations the perturbative theory for these quantities is
 finite order by order. All the integrals appearing in the theory are
 convergent both in the ultraviolet and in the infrared limits. This means
 that there is no perturbative mechanism to introduce a length scale into
 the theory of the structure functions. In turn this result indicated that
 as far as the perturbative theory is concerned there is no mechanism to
 shift the exponents $\zeta_n$ away from their K41 values.  Of course,
 nonperturbative effects may furnish such a mechanism and therefore in
 paper II \cite{95LP-c} (subtitled ``A Ladder to Anomalous Scaling") we
 turned to the analysis of nonperturbative effects. It was shown there that
 the renormalized perturbation theory for correlation functions that
 include velocity  derivatives (to second or higher power) exhibit in their
 perturbation expansion a logarithmic dependence on the  viscous scale
 $\eta$ \cite{95LL,94LL}. In this way the inner scale of turbulence appears
 explicitly in the analytic theory. The perturbative series could be
 resummed to obtain integrodifferential equations for some many-point
 objects of the theory.  These equations have also non-perturbative
 scale-invariant solutions that may be represented as power laws of
 $\eta$ to some exponents $\Delta$.  We argued \cite{95LP} that if
 $\Delta<\Delta_c$ where
 \begin{equation}
 \Delta_c=2-\zeta_2  \label{delta_c}
 \end{equation}
 (a situation referred to as the ``subcritical scenario"), then K41 scaling
 is asymptotically exact for infinite Re. In this case
 $K_{\epsilon\epsilon}(R) \sim \bar\epsilon^2 (\eta/R)^{2(4/3-\Delta)}$,
 and the exponent $\mu$ is identified with $2(4/3-\Delta)$. The
 renormalization length is then the inner length $\eta$.

 In paper II it was shown that the exponent $\Delta$ takes on exactly the
 critical value $\Delta = \Delta_c$ given by (\ref{delta_c}), and the
 subcritical scenario is not realized. In the present paper it will be
 shown that the correlation function of the energy dissipation field has
 the dependence:
 \begin{equation}
 K_{\epsilon\epsilon}(R) \sim \bar\epsilon^2
 (L/R)^{2\zeta_2-\zeta_4} (R/\eta)^
 {2(\Delta-\Delta_c)} .
 \label{Kee1}
 \end{equation}
 As a result of the criticality of $\Delta$ the inner scale $\eta$
 disappears from the correlation $K_{\epsilon\epsilon}(R)$ which finally
 takes on the form
 \begin{equation}
 K_{\epsilon\epsilon}(R) \simeq \bar\epsilon^2 (L/R)^{2\zeta_2-\zeta_4}
 \ .\label{Kee2}
 \end{equation}
 Note that this implies that for K41 scaling (in which $2\zeta_2=\zeta_4$),
 $K_{\epsilon\epsilon}(R)$ does not decay as a function of
 $R$ (i.e. the correlation is not mixing). This is hardly physical for a
 random field. On the other hand if there exists anomalous scaling in the
 sense that $\zeta_4 < 2\zeta_2$, then mixing is regained.  Thus we will
 argue that the critical scenario $\Delta = \Delta_c$ goes hand in hand
 with anomalous scaling if $K_{\epsilon\epsilon}(R)$ is mixing, and then
 $\mu$ is identified with $\zeta_4 - 2\zeta_2$,
 \begin{equation}
 \mu =\zeta_4 - 2\zeta_2 \ . \label{mu}
 \end{equation}
 This is one of the exact scaling relations that are derived below.

 The main purpose of the present paper is to elucidate the possible
 mechanisms for anomalous (non-K41) scaling of the structure functions in
 the theory of turbulence in fluids whose dynamics is described by the
 Navier-Stokes equations.  We show that the analysis of the statistical
 theory of Navier Stokes dynamics leads to the possibility of
 multiscaling with nonlinear dependence of $\zeta_n$ on $n$. For such a
 scenario we offer an evaluation of the scaling exponents $\zeta_n$ from
 first principles. We will show that the mechanisms for anomalous scaling
 are wholly non-perturbative. Order by order considerations always lead to
 K41 scaling. We show that a useful analysis of anomalous scaling can be
 developed on the basis of the so-called ``balance equations" which are
 non-perturbative.  In section 2 we derive the balance equations
 (\ref{balsc}, \ref{balvec}) by writing the equations of motion for the
 structure function $S_n(R)$ and of related quantities. These equations, in
 the stationary state, exhibit a balance between a convective (interaction)
 term which is denoted as $D_n(R)$  and a dissipative term, denoted by
 $J_n(R)$, which is the cross-correlation between the energy dissipation
 $\hat\epsilon$ and $(n-2)$ velocity differences across a scale $R$. Both
 terms are expressed as many-point correlation functions that depend on
 many coordinates, but some of these coordinates are the same. This last
 fact leads to the main strategy of this paper, which is to understand the
 fusion rules which describe the scaling of many-point correlation
 functions when some of their coordinates fuse together. When this happens
 the distance between the coordinates crosses the dissipative scale, and at
 that very moment the ultraviolet divergences are picked up. Since we
 learned in paper II how to compute these exactly, we have an important
 part of the fusion rules at hand. Following this strategy we succeed to
 compute {\it exactly} the dissipative term $J_n(R)$ and the correlation
 function $K_{\epsilon\epsilon}(R)$.

 The fusion rules are developed in sections 3,4 and 5. In section 3 we
 show that the exponent $\Delta$ of paper II is indeed the relevant
 exponent for describing pair coalescence. In section 4 we discuss
 four-point correlations and their 2-point fusion rules. This leads to the
 calculation of the correlation $K_{\epsilon\epsilon}(R)$ and the
 dissipative term $J_4$. In section 5 we discuss the fusion rules of
 $n$-point correlation functions, leading to the exact evaluation of
 $J_n(R)$.  A consequence of this calculation is the derivation of the
 dynamical scaling exponents $z_n$ describing the characteristic decay time
 $\tau_n(R)\propto R^{z_n}$ of $n$-point, $n$-time correlation functions of
 BL-velocity differences. We argue at the end of section 5 that all the
 exponents $z_n$ are independent of $n$ and that they equal $\zeta_2$.

 In section 6 we turn to the evaluation of the interaction term $D_n(R)$.
 Unfortunately we have not yet succeeded to evaluate $D_n$
 exactly. Accordingly we describe in sections 6 and 7 various possible
 evaluations of $D_n$ and the resulting implications on $\zeta_n$. It is
 shown in section 6 that the naive evaluation of $D_n$ leads to the scaling
 exponents of the $\beta$-model. We give arguments however leading to the
 conjecture that the naive evaluation of $D_n$ is an overestimate. We
 propose that there may exist a delicate cancellation in this evaluation,
 and the next order evaluation leads unequivocally to multiscaling.  In
 order to compute the scaling exponents $\zeta_n$ we need to know $D_n$
 very precisely, coefficients and all.

 Not having an exact theory for $D_n$ we resort in section 7 to modeling.
 Guided by some diagrammatics and some common sense we conjecture an
 analytic form for $D_n$ which conforms with our expectations about
 eddy-viscosity. This part of the paper is not rigorous, and the results of
 section 7 should be considered therefore as tentative.  The advantage of
 the approximation based on eddy viscosity is that we can offer analytic
 estimates of all the exponents $\zeta_n$. We discuss these estimates and
 compare them with experimental findings in Table 1 of section 7B2. Section
 8 is a summary and some discussion of the road ahead.

 Throughout the text we may refer to equations appearing in papers 0, I or
 II. When we do so we denote them as Eqs. (0--n), (I--n), (II--p) etc.
 \section{Derivation of the Balance Equations}
 Some of the most important nonperturbative constraints on the statistical
 theory of turbulence are the balance equations (also known as the moment
 equations) for the structure functions. These equations are derived in
 this section and used in later sections to deduce the scenarios for
 multiscaling in the structure functions. Before we derive the equations we
 introduce the statistical objects that appear naturally in the discussion.
 \subsection{Structure functions and related quantities}
 In this subsection we define quantities that are related to the structure
 functions $S_n(R)$. It is common in experiments to measure only the
 longitudinal structure functions \cite{MY-2}. For theoretical treatment
 these quantities are not necessarily the most convenient since they are
 not invariant to rotations. It is useful therefore to introduce some
 related objects that have simple transformation properties under rotations
 and inversions. The first one is the scalar quantity  which is appropriate
 for even orders of $S_n$. To keep in mind its scalar nature we will
 denote it as $\stackrel{o}{S}_{2m}(R)$ and define it in Eulerian terms as
 \begin{equation}
 \stackrel{o}{S}_{2m}(R) \equiv \left< |\delta{\bf u}({\bf r}_0|{\bf R},t)
 |^{2m}\right>
 \,,\quad {\bf R}\equiv {\bf r}-{\bf r}_0 \ ,
 \label{defSsc}
 \end{equation}
 where
 \begin{equation}
 \delta{\bf u}({\bf r}_0|{\bf R},t)
 \equiv{\bf u}({\bf r}_0+{\bf R},t) - {\bf u}({\bf r}_0,t)
 \label{delu}
 \end{equation}
 is a simultaneous Eulerian velocity difference. The quantity
 $\stackrel{o}{S}_{2m}(R)$ is analytic. For odd order structure functions
 we introduce a vector object ${\bf S}_{2m+1}({\bf R})$ according to
 \begin{equation}
 S^{\alpha}_{2m+1}({\bf R})\equiv \left<\delta u_{\alpha}({\bf r}_0|{\bf R},t)
 |\delta{\bf u}({\bf r}_0|{\bf R},t) |^{2m}\right> \ .	\label{defSvec}
 \end{equation}
 Here and below we will use Greek indices to indicate vector and tensor
 components, and Roman indices to indicate the order of the quantity. The
 placement of indices as subscripts or superscripts has no meaning, and is
 chosen for convenience.

 For isotropic turbulence the vector $S^{\alpha}_{2m+1}({\bf R})$ can only
 be oriented along ${\bf R}$. This allows us the introduction of  a scalar
 quantity $S_{2m+1}(R)$ which depends on the magnitude of $R$:
 \begin{equation}
 S^{\alpha}_{2m+1}({\bf R}) = {R_\alpha \over R}S_{2m+1}(R) \ . \label{S2m+1}
 \end{equation}
 Lastly we will need also the tensor objects $S^{\alpha\beta}_{2m+2}({\bf
 R}$):
 \begin{equation}
 S^{\alpha\beta}_{2m+2}({\bf R})\equiv
 \left<\delta u_{\alpha}({\bf r}_0|{\bf R},t)\delta u_{\beta}
 ({\bf r}_0|{\bf R},t)
 |\delta{\bf u}({\bf r}_0|{\bf R},t) |^{2m}\right> \ .	\label{defSten}
 \end{equation}
 Note that the objects introduced in (\ref{defSsc}, \ref{defSvec},
 \ref{defSten}) involve an arbitrary number of velocity fields but only two
 spatial points separated by $\bf R$.  We refer to them loosely as
 "two-point" correlation functions of velocity differences.  The theory
 below calls for the introduction of three-point functions as well.  It is
 best to define them using the four-point functions
 \FL
 \begin{eqnarray}
 && T^{\alpha\beta}_{2m+2}({\bf R}_1,{\bf R}_2,{\bf R})
 \equiv \Big\langle \delta  u_{\alpha}({\bf r}_0|{\bf R}_1,t)
 \delta u_{\beta}({\bf r}_0|{\bf R}_2,t)
  \label{T2} \\
 &&\qquad \qquad \times
 |\delta{\bf u}({\bf r}_0|{\bf R},t) |^{2m}\Big\rangle \ ,
 \nonumber \\
 && T^{\alpha\beta\gamma}_{2m+3}({\bf R}_1,{\bf R}_2,{\bf R})
 \equiv \Big\langle \delta u_{\alpha}({\bf r}_0|{\bf R}_1,t)
 \delta u_{\beta}({\bf r}_0|{\bf R}_2,t)
 \label{T3} \\  &&\qquad  \qquad\times
 \delta u_{\gamma}({\bf r}_0|{\bf R},t)|\delta{\bf u}({\bf
 r}_0|{\bf R},t) |^{2m}\Big\rangle \ ,
 \nonumber \\
 && T^{\alpha\beta\gamma\delta}_{2m+4}({\bf R}_1,{\bf R}_2,{\bf R})
 \equiv \Big\langle \delta u_{\alpha}({\bf r}_0|{\bf R}_1,t)
 \delta u_{\beta}({\bf r}_0|{\bf R}_2,t)
 \label{T4}
  \\  &&\qquad  \qquad \times
 \delta u_{\gamma}({\bf r}_0|{\bf R},t)\delta
 u_{\delta}({\bf r}_0|{\bf R},t)|\delta{\bf u} ({\bf r}_0|{\bf R},t)
 |^{2m}\Big\rangle \ .
 \nonumber
 \end{eqnarray}
 We will see that the theory produces expressions invloving these
 quantities with two of the arguments being identical (i.e. ${\bf R}_1={\bf
 R}_2$, etc.) This fact will lead to the study of fusion rules in later
 sections.
 \subsection{The Balance Equations}
 In this subsection we derive the balance equation that relates structure
 funtions to correlation functions involving the dissipation field. We
 start by deriving some equations of motion for the quantities defined
 above.
 \subsubsection{Equations of Motion}
 Our starting point is the Navier-Stokes equations for an incompressible
 fluid:
 \begin{eqnarray}
  {\partial{\bf u}({\bf r},t)\over\partial t}&+&[{\bf u}({\bf r},t)
 \cdot\nabla ]{\bf u}({\bf r},t)
  -\nu\nabla^2{\bf u}({\bf r},t)-\nabla\,p({\bf r},t)
 \nonumber \\
 &&={\bf f}({\bf r},t)\,,\qquad
 \nabla\cdot{\bf u}=0 \ .
  \label{A3}
  \end{eqnarray}
 For simplicity we will choose the forcing such that $\nabla\cdot{\bf f} =
 0$.  This equation can be rewritten in terms of the Belinicher-L'vov
 velocities ${\bf v}({\bf r}_0,{\bf r},t)$ defined by Eq. (I--2.2) as
 follows:
 \begin{eqnarray}
 {\partial{\bf v}({\bf r}_0|{\bf r},t)\over\partial t}&+&[{\bf w}({\bf
  r}_0|{\bf r},t) \cdot\nabla ]{\bf w}({\bf r}_0|{\bf r},t)
  -\nu\nabla^2{\bf w}({\bf r}_0|{\bf r},t)
 \nonumber \\
 &-&\nabla\,\tilde p({\bf r}_0|{\bf
  r},t)= \tilde {\bf f}({\bf r}_0|{\bf r},t) \ , \label{BL}
 \end{eqnarray}
 where we also have the incompressibility condition $\nabla\cdot{\bf w}=0$.
 Here ${\bf w}$ is the BL velocity difference
 \begin{equation}
 {\bf w}({\bf r}_0|{\bf r},t)\equiv{\bf v}({\bf r}_0|{\bf r},t)-{\bf
 v}({\bf r}_0|{\bf r}_0,t) \ , \label{delw}
 \end{equation}
 and $\tilde p({\bf r}_0|{\bf r},t)$, $\tilde {\bf f}({\bf r}_0|{\bf r},t)$
 are BL-transformed pressure and forcing (for more detail, see Section 3A of
 paper I).  Applying the transverse projector $\OP$ this equation takes on
 the form
 \begin{eqnarray}
 {\partial{\bf v}({\bf r}_0|{\bf r},t)\over\partial t} &+&\OP[{\bf w}({\bf
 r}_0|{\bf r},t)\cdot\nabla ]{\bf w}({\bf r}_0|{\bf r},t)
 \label{NSpro} \\
 &-&\nu\nabla^2{\bf  w}({\bf r}_0|{\bf r},t)=\tilde {\bf f}({\bf r}_0|{\bf
 r},t) \ .  \nonumber
 \end{eqnarray}
 The application of ${\raisebox{.2ex}{$\stackrel{\leftrightarrow}{\bf P}$}
 }$ to any given vector field ${\bf a}( {\bf r})$ is non-local, and
 has the form:
 \begin{equation}
 \lbrack\raisebox{.2ex}{$\stackrel{\leftrightarrow}{\bf P}$}
  {\bf a}(
 {\bf r})\rbrack_\alpha =\int d   {\bf r}'   P_{\alpha\beta}(
 {\bf r}-{\bf r}')a_\beta(   {\bf r}'),
 \label{b2}
 \end{equation}
 where $P_{\alpha\beta}(   {\bf r}-   {\bf r}')$ is
 \begin{eqnarray}
 P_{\alpha\beta}({\bf r}&-&{\bf r}')
 =\delta_{\alpha\beta}\delta({\bf r}-{\bf r}' )
 \label{b5} \\
 &-&{1\over 4\pi}\left[{\delta_{\alpha\beta}
 \over \vert{\bf r}-{\bf r}' \vert^3}
 -3{(r_\alpha-r'_\alpha)(r_\beta-r'_\beta)
 \over\vert{\bf r}-{\bf r}'\vert^5}\right] \ .
 \nonumber
 \end{eqnarray}
 Next we consider Eq. (\ref{NSpro}) at two spatial points ${\bf r}$
 and ${\bf r} _0$. Subtracting the two equations we get
 \FL
 \begin{eqnarray}
 &&{\partial{w_\alpha ({\bf r}_0|{\bf r},t)}\over\partial t}+\int d{\bf r}'
 [P_{\alpha\beta}({\bf r}-{\bf r}') - P_{\alpha\beta}({\bf r}_0-{\bf r}')]
 \nonumber \\
 &&\qquad\qquad \times w_\gamma({\bf r}_0|{\bf r}',t)(\nabla_{r'})_\gamma
  w_\beta({\bf r}_0|{\bf r}',t) \label{NS2} \\
  &&= \nu[\nabla_r^2+\nabla_{r_0}^2]w_\alpha ({\bf r}_0|{\bf r},t)+\tilde
 f_\alpha({\bf r}_0|{\bf r},t)- \tilde f_\alpha({\bf r}_0|{\bf r}_0,t)\ ,
  \nonumber
 \end{eqnarray}
 where $(\nabla_{r})_\gamma \equiv \partial/\partial r_\gamma$.  Introduce
 now the shorthand notation $w^2$ and $w^{2m}$ in situations in
 which the arguments are $({\bf r}_0|{\bf r},t)$:
 \begin{equation}
 | {\bf w}({\bf r}_0|{\bf r},t) |^2 \equiv w ^2 \ ,
 \quad |{\bf w}({\bf r}_0|{\bf r},t)|^{2m}\equiv  w^{2m}		\ .
 \label{defs}
 \end{equation}
 When other arguments appear they will be displayed explicitly.  In term of
 these quantities the structure functions (\ref{defSsc})-(\ref{defSvec})
 are written as
 \begin{eqnarray}
 \stackrel{o}{S}_{2m}(R) &\equiv& \left<w^{2m}\right> ,	\label{eqSsc} \\
 {\bf S} _{2m+1}(R) &\equiv& \left< {\bf w} w^{2m}\right> \ .
 \label{eqSvec}
 \end{eqnarray}
 where again ${\bf R} = {\bf r}-{\bf r}_0$ and  ${\bf w}$ without arguments
 means ${\bf w}({\bf r}_0|{\bf r},t)$. Next observe that
 \begin{equation}
 {\partial w^{2m}\over{\partial t}} = 2m w^{2(m-1)} {\bf w}
 \cdot{\partial{\bf w}\over \partial t} \ .	\label{eqv2m}
 \end{equation}
 Evaluate now the scalar product obtained from Eq. (\ref{NS2}) by
 multiplying it on the left by $2mw^{2(m-1)}{\bf w}$.  Using Eq.
 (\ref{eqv2m}) this is written exactly as
 \begin{equation}
 {\partial \stackrel{o}{S}_{2m}(R)\over \partial t} + D_{2m}(R)
 = J_{2m}(R) + Q_{2m}(R) \ ,
 \label{balancesc}
 \end{equation}
 where we denoted
 \begin{eqnarray}
 D_{2m}(R) &\equiv& 2m\int d{\bf r}_1 P_{\alpha\beta}({\bf r}_1)
 {\partial\over\partial r_{1\gamma}}
 \Big\langle w^{2(m-1)} w_{\alpha}
 \nonumber \\
 &&\times
 \Big[w_\gamma({\bf r}_0|{\bf r}+{\bf r}_1,t)w_\beta
 ({\bf r}_0|{\bf r}+{\bf r}_1,t) \nonumber \\
 && -w_\gamma({\bf r}_0|{\bf r}_0+{\bf r}_1,t)
 w_\beta({\bf r}_0|{\bf r}_0+{\bf r}_1,t)\Big]
 \Big\rangle \ , \label{D2m} \\
 J_{2m}(R) &\equiv& 2m \nu \Big\langle w^{2(m-1)}
 w_{\alpha}\Big[\nabla_r^2+\nabla_{r_0}^2\Big]
 w_\alpha \Big\rangle \ ,
 \label{J2m}	\\
 Q_{2m}(R) &\equiv& 2m \Big\langle w^{2(m-1)}
 w_{\alpha}\Big[f_\alpha({\bf r}_0|{\bf r},t)-
  f_\alpha({\bf r}_0|{\bf r}_0,t)\Big]\Big\rangle \ .
 \nonumber \\
 \label{Q2m}
 \end{eqnarray}
 In deriving the equation for $D_{2m}$ we used the incompressibility
 condition (which is not available in the Burgers' equation) and performed
 changes of variables in the integrals according to ${\bf r}'-{\bf r}={\bf
 r}_1$ and ${\bf r}'-{\bf r}_0={\bf r}_1$. In the stationary state
 $\partial\stackrel{o}{S}_{2m}(R)/ \partial t = 0$, and we can write the
 scalar balance equation
 \begin{equation}
 D_{2m}(R)  = J_{2m}(R) + Q_{2m}(R) \ . \label{balsc}
 \end{equation}
 Next we need to derive vectorial balance equations for ${\bf S}_{2m+1}$.
 Repeating similar steps we end up with
 \begin{equation}
 {\bf D}_{2m+1}({\bf R})
 = {\bf J}_{2m+1}({\bf R}) + {\bf Q}_{2m+1}({\bf R}) \ ,\label{balvec}
 \end{equation}
 where
 \FL
 \begin{eqnarray}
 && D^\delta_{2m+1}({\bf R})=2m\int d{\bf r}_1 P_{\alpha\beta}({\bf r}_1)
 {\partial\over\partial r_{1\gamma}}
 \Big\langle w^{2(m-1)}w_\delta w_{\alpha}
 \nonumber \\
 &&\qquad\qquad\times
 \Big[w_\gamma({\bf r}_0|{\bf r}+{\bf r}_1,t)w_\beta
 ({\bf r}_0|{\bf r}+{\bf r}_1,t)
 \nonumber \\
 &&\qquad\qquad-w_\gamma({\bf r}_0|{\bf r}_0+{\bf r}_1,t)
 w_\beta({\bf r}_0|{\bf r}_0+{\bf r}_1,t)\Big]   \Big\rangle
 \nonumber \\
 \!&+&\!\! \int \! d{\bf r}_1 P_{\delta\beta}({\bf r}_1)
 {\partial\over\partial r_{1\gamma}}
 \Big\langle w^{2m} \big[w_\gamma({\bf r}_0|{\bf r}+{\bf r}_1,t)w_\beta
 ({\bf r}_0|{\bf r}+{\bf r}_1,t)
 \nonumber \\
 &&\qquad-w_\gamma({\bf r}_0|{\bf r}_0+{\bf r}_1,t)
 w_\beta({\bf r}_0|{\bf r}_0+{\bf r}_1,t)\big]
 \Big\rangle\ ,
 \label{Dalph} \\
 &&J^\delta_{2m+1}({\bf R})  = 2m \nu \Big
 \langle w^{2(m-1)}w_\delta w_{\alpha}[\nabla_r^2+
 \nabla_{r_0}^2] w_\alpha \Big\rangle
 \nonumber \\
 &&\qquad\qquad+\nu \Big\langle w^{2m}[\nabla_r^2
 +\nabla_{r_0}^2] w_\delta \Big\rangle \ ,
 \label{Jalph} \\
 &&\!\!\!\!\!\! Q^\delta_{2m+1}({\bf R})\! = \!
 2m \Big\langle w^{2(m-1)}w_\delta w_{\alpha}
 [f_\alpha({\bf r}_0|{\bf r},t)\!
 - \! f_\alpha({\bf r}_0|{\bf r}_0,t)]\Big\rangle
 \nonumber \\
 &&\qquad\qquad+\Big\langle w^{2m} [f_\delta({\bf r}_0|{\bf r},t)-
  f_\delta({\bf r}_0|{\bf r}_0,t)]\Big\rangle \ .
 \label{Qalph}
 \end{eqnarray}
 For isotropic turbulence all these vector quantities have obvious scalar
 counterparts, similar to the one introduced in (\ref{S2m+1}). For example,
 \begin{eqnarray}
 D^{\alpha}_{2m+1}({\bf R}) &=& {R_\alpha \over R}D_{2m+1}(R) \ .
 \label{D2m+1} \ , \\
 J^{\alpha}_{2m+1}({\bf R}) &=& {R_\alpha \over R}J_{2m+1}(R) \ .
 \label{J2m+1}
 \end{eqnarray}
 With these definitions we can rewrite the balance equation (\ref{balvec})
 in the form
 \begin{equation}
 D_{2m+1}(R)  = J_{2m+1}(R) + Q_{2m+1}(R) \ . \label{balvec2}
 \end{equation}
 Together with (\ref{balsc}) we can finally write for $n$ odd or even
 \begin{equation}
 D_n(R)  = J_n(R) + Q_n(R) \ . \label{balfinal}
 \end{equation}
 We discuss each term in the balance equation in the following subsections.
 \subsection{The Terms of the Balance Equation}
 \subsubsection{The Interaction Term}
 To understand how to evaluate $D_{2m}(R)$ we begin the discussion by
 neglecting the effect of the pressure term in the Navier-Stokes equation.
 This is equivalent to replacing the kernel of the projection operator
 $P_{\alpha\beta}(r)$ by a the unit operator
 $\delta_{\alpha\beta}\delta({\bf r})$. The effect of this replacement of
 kernels is to change $D_{2m}(R)$ to
 \begin{equation}
 D_{2m}(R) = 2m\left<w^{2(m-1)} w_{\alpha} w_{\gamma}
 {\partial w_{\alpha}\over \partial r_{\gamma}}\right>
 \ . \label{D2mint}
 \end{equation}
 This may be rewritten as
 \begin{eqnarray}
 D_{2m}(R)& =& {\partial \over \partial r_{\gamma}}\left<w_\gamma w^{2m}
 \right>=  {\partial \over \partial R_{\gamma}}S_{2m+1}^\gamma ({\bf R})
 \label{D2mnop}   \\
 &&(\rm neglecting~pressure) \ .
 \nonumber
 \end{eqnarray}
 This expression is relevant for the 1-dimensional Burgers equation,
 since there is no pressure in that case. On the other hand the Burgers
 equation describes a compressible flow, and this fact introduces the
 coefficient $(n+1)/n$ in the RHS of this expression.  For the
 Navier-Stokes equations we can only use the expression for a rough
 evaluation of $D_{2m}$. Notwithstanding, it is exact for the case $m=1$,
 where it takes the form
 \begin{equation}
 D_2(R) =  {\partial\over R_\alpha} S^\alpha_3 ({\bf R})\ . \label{D2}
 \end{equation}
 The deep reason for this simplification in the case $m=1$ is the
 conservation law (in the absence of viscosity) of the quadratic invariant
 of the BL equation of motion (\ref{BL}), i.e. $ \int d{\bf r} {\bf
 w}^2[{\bf r}_0|{\bf r},t])$.  $D_2(R)$ is simply the rate of change of
 ${\bf w}^2[{\bf r}_0|{\bf R},t]$ and therefore must be a pure divergence.
 None of the higher powers of ${\bf w}[{\bf r}_0|{\bf R},t]$ forms an
 invariant, and as a consequence none of the higher orders $D_n(R)$ can be
 written as a pure divergence. On the other hand we see from Eq. (\ref{D2m})
 that with the full kernel we have two terms that can be expressed in terms
 of the $T$ tensors:
 \begin{eqnarray}
 && D_{2m}(R) = 2 m\int d{\bf r}_1  P_{\alpha\beta}({\bf r}_1)
  \label{D2mfinal}  \\
 &\times&{\partial\over \partial
 r_{1\gamma}}\big[T_{2m+1}^{\alpha\beta\gamma}({\bf R}+{\bf r}_1,
 {\bf R}+{\bf r}_1,{\bf R})-
 T_{2m+1}^{\alpha\beta\gamma}({\bf r}_1,{\bf r}_1,{\bf R})\big] \ .
 \nonumber
 \end{eqnarray}
 This is the final form of $D_{2m}(R)$.

 Following the same steps of analysis we find the final expression for
 ${\bf D}_{2m+1}({\bf R})$.
 \begin{eqnarray}
 && D^\delta_{2m+1}({\bf R})
 =2m\int d{\bf r}_1 P_{\alpha\beta}({\bf r}_1)
 \label{D2vecfinal}\\
 &\times&
 {\partial\over \partial r_{1\gamma}}
 \Big\{T^{\alpha\beta\gamma\delta}_{2m+2}({\bf R}+{\bf r}_1,{\bf R}
 +{\bf r}_1,{\bf R}) - T^{\alpha\beta\gamma\delta}_{2m+2}({\bf r}_1,{\bf
 r}_1,{\bf R})\Big\}
 \nonumber\\
 &&\ \ +\int d{\bf r}_1 P_{\delta\beta}({\bf r}_1) {\partial\over \partial
 r_{1\gamma}} \Big\{T^{\beta\gamma}_{2m+2}({\bf R}+{\bf r}_1,{\bf
 R}+{\bf r}_1,{\bf R})
 \nonumber\\
 &&\qquad\qquad\qquad - T^{\beta\gamma}_{2m+2}({\bf r}_1,{\bf r}_1,{\bf
 R})\Big\} \ .
 \nonumber
 \end{eqnarray}
 The naive evaluation of every term in Eq. (\ref{D2mfinal}) is given by
 (\ref{D2mnop}).  However this evaluation is only acceptable under two
 assumptions: (i) that the integral converges, and (ii) that there are no
 cancellations between the terms with opposite signs in
 Eq. (\ref{D2mfinal}). The analysis of convergence requires understanding
 the asymptotic properties of the $T$ correlators. Similar properties will
 determine the	evaluation of $J_{2m}$ as will be seen next. For that
 reason we will devote the next Sections to questions of asymptotics and
 fusion rules for $n$-point correalation function of velocity
 differences. We will learn that the integral in Eq. (\ref{D2mfinal}) indeed
 converges, but on the other hand cancellations are not excluded and are
 the possible source of multiscaling in turbulence. In fact, such a
 cancellation is the only mechanism of multiscaling that we can identify.
 \subsubsection{The Dissipation Term}
 Starting with (\ref{J2m}) we first use the fact that for simultaneous
 correlations we can use the Eulerian differences $\delta {\bf u}$ instead
 of the BL-velocity differences.  Secondly we use the symmetry with respect
 to exchange of ${\bf r}$ and ${\bf r}_0$ to write the expression with twice
 the Laplacian with respect to ${\bf r}_0$ only, instead of the sum of
 Laplacians. Lastly we use translational invariance to move one of the
 gradients around. We find
 \begin{eqnarray}
 J_{2m}(R)&\equiv& -4m\nu\Bigg\{ \left<|\delta {\bf u}({\bf r}_0|{\bf
 R})|^{2(m-1)} \left[{\partial\over\partial r_{0\beta}} u_\alpha({\bf
 r}_0)\right]^2\right >
 \nonumber \\
 + 2(m&-&1) \Bigg<|\delta {\bf u}({\bf r}_0|{\bf R})|^{2(m-2)}
 \delta u_\alpha({\bf r}_0|{\bf R}) \delta u_\gamma({\bf r}_0|{\bf R})
 \nonumber\\
 &&\times {\partial\over\partial r_{0\beta}}
 u_\gamma({\bf r}_0) {\partial\over\partial r_{0\beta}} u_\alpha({\bf
 r}_0)\Bigg> \Bigg\}\ .
  \label{J2mint1}
 \end{eqnarray}
 To relate this quantity to the tensors $T$ defined in Eqs.
 (\ref{T2}-\ref{T4}) we write the derivative as the limit of velocity
 differences:
 \begin{equation}
 {\partial\over\partial r_{0\beta}} u_\alpha({\bf r}_0) \equiv
 \lim_{d_\beta \to 0} {u_\alpha({\bf r}_0+d_\beta {\bf
 e}_\beta)-u_\alpha({\bf r}_0)\over d_\beta} \ , \label{deriv}
 \end{equation}
 where ${\bf e}_\beta$ is the unit vector in the $\beta$ direction.
 With this in mind the expression for $J_{2m}$ is
 \begin{eqnarray}
 J_{2m}(R) &=& - 4m\nu\lim_{d_\beta \to 0}{1\over d^2_\beta}
 \Big[T^{\alpha\alpha}_{2m}(d_\beta{\bf e}_\beta,d_\beta{\bf e}_\beta,{\bf
 R}) \nonumber\\ &&+ 2(m-1)T^{\alpha\gamma\alpha\gamma}_{2m}(d_\beta{\bf
 e}_\beta,d_\beta{\bf e}_\beta,{\bf R})\Big] \ .
 \label{J2mfinal}
 \end{eqnarray}
 One should understand that a summation over $\alpha,\beta$ and $\gamma$ is
 implied as usual.  Note that for $d_\beta \ll \eta$ the velocity field is
 expected to be smooth, and both $T$ functions become proportional to
 $d_\beta^2$, cancelling this factor in the denominator.  We will not have
 a dependence of the limit on the direction of the vector ${\bf d}\equiv
 d_\beta {\bf e}_\beta$.

 It can be seen that the evaluation of $J_{2m}(R)$ again requires
 elucidation of the asymptotic properties of the correlations functions
 including the rules of coalescence of groups of points. In paper I we
 proved the property of locality that means that the coalescence itself is
 regular. However, we have here differential operators on coalescing points
 and we need to consider not only leading behaviours but also next order
 properties. This is done in the section 5.

 The quantities $J^\alpha_{2m+1}({\bf R})$ can be expressed in terms of the
 third and fifth rank $T$ tensors, but we avoid displaying the result since
 we do not use it explicitly.

 Lastly we need to discuss the forcing terms $Q_{2m}(R)$ and $Q^\alpha_
 {2m+1}({\bf R})$. This is the topic of the next subsection.
 \subsubsection{The Forcing Term}
 In this subsection we show that the forcing term can be neglected in the
 inertial range of scales. The reader who finds this statement believable
 can skip this subsection.  To discuss the forcing terms it is useful to
 introduce a higher order Green's function according to
 \begin{equation}
 {\cal G}^{\alpha\beta}_{2m-1,1}({\bf r}_0|x,x') \equiv \left<{\delta
 w^{2(m-1)}w_\alpha\over \delta \tilde f_\beta({\bf r}_0|x')} \right> \ .
 \label{a2}
 \end{equation}
 We remind the reader that $x=({\bf r} ,t)$ and that according to our
 convention we can omit the argument $({\bf r}_0|x)$ in the function ${\bf
 w}$.

 When the statistics of the random forcing is Gaussian we can write
 $Q_{2m}(R)$ of Eq. (\ref{J2m}) as
 \begin{eqnarray}
 Q_{2m}(R) &=&\int d{\bf r} ' dt'{\cal G}^{\alpha\beta}_{2m-1,1}({\bf
 r}_0|x,x')\Big[D_{\alpha\beta}({\bf r}_0| {\bf r}',{\bf r},t'-t)
 \nonumber \\
 && -D_{\alpha\beta}({\bf r}_0|{\bf r}',{\bf r}_0,t'-t)\Big] \ .
 \label{a1}
 \end{eqnarray}
 Here, as usual, ${\bf R}={\bf r}-{\bf r}_0$ and $D_{\alpha\beta}({\bf
 r}_0|x,x')$ is the correlator of $\tilde f_{\alpha(x)}$ and $\tilde
 f_{\beta}(x)$.

 We are going to estimate $Q_{2m}(R)$ by taking first a delta-correlated
 forcing,
 \begin{equation}
 D_{\alpha\beta}({\bf r}_0|x,x') = D_{\alpha\beta}({\bf r}-{\bf r}
 ')\delta(t-t') \ . \label{a3}
 \end{equation}
 The consideration of the effect of finite correlation time is deferred to
 the end.  Using (\ref{a3}) Eq. (\ref{a1}) simplifies to
 \begin{eqnarray}
 Q_{2m}(R) &=&\int d{\bf r} '{\cal G}^{\alpha\beta}_{2m-1,1}({\bf r}_0|{\bf
 r},{\bf r} ',0)\Big[D_{\alpha\beta} ({\bf r}'-{\bf r})
 \nonumber\\
 &-&D_{\alpha\beta}({\bf r}'-{\bf r}_0)\Big] \ . \label{a4}
 \end{eqnarray}
 Because of the delta functions we have lost the time integration and we have
 the zero-time Green's function, which is not dressed by the interaction
 (and see paper I for an explicit demonstration). Using the chain rule of
 differentiation in (\ref{a2}) we get
 \begin{eqnarray}
 {\cal G}^{\alpha\beta}_{2m-1,1}({\bf r}_0|x,x') =\left<w^{2(m-1)}{\delta
 w_\alpha\over \delta \tilde f_\beta({\bf r}_0|x')} \right> &&
 \label{intermed}\\
 + 2(m-1)\left<w^{2(m-2)}w_\alpha w_\gamma {\delta w_\gamma\over \delta
 \tilde f_\beta ({\bf r}_0|x')} \right> && \ .
 \nonumber
 \end{eqnarray}
 Next we recall that at time $t=0$ the unaveraged response $\delta
 w/\delta\tilde f$ is uncorrelated with the velocity field since any
 interaction involves a vertex with time integration between $0$ and $t$.
 Accordingly we can decouple the response in (\ref{intermed}) and write
 \begin{eqnarray}
 {\cal G}^{\alpha\beta}_{2m-1,1}({\bf r}_0|x,x')&=& {\cal
 G}^0_{\alpha\beta}({\bf r}_0|{\bf r},{\bf r}') \stackrel{o}{S}_{2(m-1)}(R)
 \label{a5}  \\
 &&+2(m-1){\cal G}^0_{\gamma\beta}({\bf r}_0|{\bf
 r},{\bf r}')S^{\alpha\gamma}_ {2(m-1)}(R) \ ,
 \nonumber
 \end{eqnarray}
 where ${\cal G}^0$ is the bare Green's function which was defined and
 computed in paper I sections 2,3.  We proceed to evaluate the order of
 magnitude of $Q_{2m}$ by substituting the last equation in (\ref{a4}).
 Suppressing vector indices we evaluate
 \begin{equation}
 Q_{2m}(R) \simeq
 m(2m-1)S_{2(m-1)}(R) Q_2(R) \ ,
 \label{a6}
 \end{equation}
 where
 \begin{equation}
 Q_2(R) =\int d{\bf r} '{\cal G}^0({\bf r}_0|{\bf r},{\bf
 r} ',0)\Big[D({\bf r}'-{\bf r})-D({\bf r}'-{\bf r}_0)\Big] \ .
 \label{a7}
 \end{equation}
 We can estimate $Q_2$ by recalling that $D(R)$ has a characteristic scale
 $L$ (which is the outer scale of turbulence). For $R \ll L$ we expand
 $D(R)$ in $R$:
 \begin{equation} D(R) = D(0) + D^{\prime\prime}R^2/2 \ ,
 \qquad D^{\prime\prime}\simeq D(0)/L^2 \ .
 \label{expD}
 \end{equation}
 In its turn $D(0)$ is about the rate of energy input, and for stationary
 turbulence $D(0)\simeq \bar \epsilon$, the rate of energy dissipation.
 Finally we estimate
 \begin{equation}
 D(R)-D(0) \simeq \bar \epsilon  (R/L)^2 \ .
 \label{a8}
 \end{equation}
 Together with the evaluation of $G^0$ that was presented in paper I
 Section 3 B we evaluate $Q_2$ as
 \begin{equation}
 Q_2(R) \sim {\bar\epsilon\over L^2}\int d{\bf r}'{|{\bf R}-{\bf
 r}'|^2-r'^2\over |{\bf R}-{\bf r}'|^3- r'^3} \ .  \label{a9}
 \end{equation}
 The integral can be evaluated by introducing the ultraviolet cutoff at the
 viscous scale $\eta$:
 \begin{equation}
 Q_2 \sim \bar\epsilon \left({R\over L}\right)^2
 \ln\left({R\over\eta}\right) \ .
 \label{a10}
 \end{equation}
 Substituting in (\ref{a6}) and omitting numerical factors we end up with
 \begin{equation}
 Q_{2m}(R) \sim \bar\epsilon \left({R\over L}\right)^2 S_{2(m-1)}
 (R)\ln\left({R\over\eta}\right)  \ .
 \label{a11}
 \end{equation}
 We are now in a position to compare $Q_{2m}$ with the interaction term
 $D_{2m}$ given in (\ref{D2mint}). For K41 scaling and delta-correlated
 forcing we have
 \begin{equation} {Q_{2m}\over D_{2m}} \sim \left({R\over
 L}\right)^2 \ln\left({R\over\eta}\right) \ .  \label{a12}
 \end{equation}
 It should be noted that the logarithm in this equation is a direct result
 of our use of delta correlated forcing. This choice  brought in the
 zero-time Green's function with its inverse cubic $R$ dependence, which
 led to the logarithm. Any realistic forcing would be integrated against
 the finite time response which is less singular than the zero time
 function.  We thus assert at this point that a more detailed analysis
 should reveal that $Q_{2m}$ is negligible compared to $D_{2m}$ in the core
 of the inertial interval. The power law $(R/L)^2$ for the ratio of these
 quantities is appropriate for K41 scaling, and the exponent may be
 slightly different for non-K41 exponents. At any rate we are satisfied
 that for $R\ll L$ the forcing term is negligible compared to the transfer
 term.
 \section{The Anomalous Exponent Associated with Pair Coalescence}
 In this section we begin the exploration of the ultraviolet properties of
 $n$-point correlation functions with the aim of computing $J_n$ and $D_n$
 of the previous section. To achieve this we examine first a second-order
 Green's function of the type introduced in paper II and whose ultraviolet
 properties are characterized by an anomalous exponent $\Delta$. We will
 prove that these ultraviolet properties are shared by the second order
 structure function, and that $\Delta$ attains	the critical value
 $\Delta=2-\zeta_2$. Although this statement was made already in paper II
 section 5.B, we discuss it here in full detail because of its crucial
 importance for the structure of the theory of turbulence, and in
 particular in determining also the ultraviolet properties of the
 quantities appearing in the balance equations (\ref{balsc}) and
 (\ref{balvec}). We begin with the 4-point Green's functions. The analysis
 of this section will allow us already to evaluate the correlation function
 of the dissipation field (\ref{Kee}) and to derive the important scaling
 relation (\ref{mu}). The strategy of this section is to first introduce
 the needed many-point Green's functions, then to examine and classify their
 diagrammatic representation, and lastly to resum the diagrammatic series
 to obtain exact integral equations for these quantities. The exact
 equations will allow us to find non-perturbative solutions.
 \subsection{Many-Point Green's Functions}
 \subsubsection{Definitions}
 Consider the nonlinear Green's functions  ${\cal G}_{m,n}$ which are the
 response of the product of $m$ BL-velocity differences to $n$
 perturbations. In particular
 \FL
 \begin{eqnarray}
 {\cal G}_{2,1}^{\alpha\beta\gamma}({\bf r}_0|x_1,x_2,x_3)&=&
 \left<{\delta [w_\alpha({\bf r}_0|x_1)
   w_\beta({\bf r}_0|x_2)]\over \delta h_{\gamma}(x_3)}
 \right> \ , \label{newG21} \\
 {\cal G}_{1,2}^{\alpha\beta\gamma}({\bf r}_0|x_1,x_2,x_3)&=&
 \left<{\delta^2 w_\alpha({\bf r}_0|x_1)
  \over  \delta h_\beta({\bf r}_0|x_2)\delta h_{\gamma}(x_3)}
 \right> \ , \label{newG12}
 \end{eqnarray}
 \FL
 \begin{eqnarray}
 &&
 {\cal G}_{2,2}^{\alpha\beta\gamma\delta}({\bf r}_0|x_1,x_2,x_3,x_4) =
 \left<{\delta^2 [w_\alpha({\bf r}_0|x_1) w_\beta({\bf r}_0|x_2)]\over
   \delta h_{\gamma}(x_3)\delta h_{\delta}(x_4)} \right> \ ,
 \nonumber\\
 && \label{newG22} \\
 &&
 {\cal G}_{3,1}^{\alpha\beta\gamma\delta}({\bf r}_0|x_1,x_2,x_3,x_4)
 \nonumber\\
 &&\qquad\qquad =\left<{\delta [w_\alpha({\bf r}_0|x_1)
   w_\beta({\bf r}_0|x_2) w_{\gamma}(x_3)]\over \delta h_{\delta}(x_4)}
 \right> \ . \label{newG31}
 \end{eqnarray}
 Note that these are different from the nonlinear Greens' function $G_2$
 that was dealt with extensively in paper II,
 \begin{equation}
 G_2^{\alpha\beta\gamma\delta}({\bf r}_0|x_1,x_2,x_3,x_4)
  =  \left<{\delta w_\alpha({\bf r}_0|x_1)
   \over \delta h_{\gamma}(x_3)}{\delta w_\beta({\bf r}_0|x_2)
 \over \delta h_{\delta}(x_4)}
 \right> \ . \label{oldG}
 \end{equation}
 One can see that the relation between these 4-point Green's functions follows
 from the chain rule of differentiation and is
 \begin{eqnarray}
 &&{\cal G}_{2,2}^{\alpha\beta\gamma\delta}({\bf r}_0|x_1,x_2,x_3,x_4)
  = G_2^{\alpha\beta\gamma\delta}({\bf r}_0|x_1,x_2,x_3,x_4)
 \nonumber \\
 &+&G  _2^{\alpha\beta\delta\gamma}({\bf r}_0|x_1,x_2,x_4,x_3)
 +\left<{ w_\alpha({\bf r}_0|x_1) \delta^2 w_\beta({\bf r}_0|x_2)\over
   \delta h_{\gamma}(x_3) \delta h_{\delta}(x_4)} \right>
 \nonumber \\
 &+&\left<{ w_\beta({\bf r}_0|x_2) \delta^2 w_\alpha({\bf r}_0|x_1)\over
 \delta h_{\gamma}(x_3)\delta h_{\delta}(x_4)} \right> \ .
 \label{relation}
 \end{eqnarray}
 In order to analyze the properties of the functions ${\cal G}_{m,n}$ we
 will explore their diagrammatic representation.
 \subsubsection{Diagrammatic Representation}
 In section 3 A of paper II it was explained in detail how to produce the
 diagrammatic representation of $G_2$. Very similar steps lead to the
 diagrams for ${\cal G}_{2,2}$, which are shown in Fig. \ref{95LP-01}. Of
 course, due to (\ref{relation}) all the diagrams of $G_2$ appear also in
 the representation of ${\cal G}_{2,2}$. These common diagrams all have
 two principal paths made of Green's functions as explained in paper II.
 In ${\cal G}_{2,2}$ there are contributions coming from the last two
 terms in (\ref{relation}). To understand the structure of these new
 diagram note that every diagram representing $\delta^2w_1/\delta
 h_3\delta h_4$ has a principal path of Green's functions that starts with
 coordinate $x_1$ or $x_2$ which splits at some point into two principal
 paths made of Green's functions ending up with the coordinates $x_3,x_4$.
 The field $w_2$ itself is another tree of the type appearing in Fig.3 of
 paper II. Upon multiplying these trees and averaging over the Gaussian
 ensemble of forces we get diagrams that are ready to be resummed. One set
 of diagrams that appear are those that have a bridge made of a single
 Green's function connecting the ``left" part of the diagram (coordinates
 ${\bf r}_1,{\bf r}_2$ to the ``right "part of the diagram (coordinates
 ${\bf r}_3,{\bf r}_4$). By necessity this Green's function belongs to the
 principal path of the diagram which is made of Green's functions. We note
 that a bridge made of a single propagator may appear only once. A second
 singly propagator bridge makes the diagram ``one-eddy reducible", and
 such diagrams have already been line-resummed; see paper 0 for more
details.  Every decoration to the left of the bridge may be resummed into
 a dressed vertex on the left, and every decoration on the right of the
 bridge may be resummed into a dressed vertex on the right. The fully
 resummed series of weakly linked (via one-propagator bridge) diagrams
 which we denote as ${\cal G}^{\rm wl}_{2,2}$ contains exactly the two
 diagrams that are shown in Fig. \ref{95LP-01}, panels b-d.

 The reader's attention should be drawn to the weakly linked diagrams which
 have appeared here for the first time. It will turn out that these
 diagrams play a very important role in the mechanism that we propose for
 anomalous scaling in turbulence. For that reason we pause for a moment to
 discuss the topological structure of these diagrams in more detail.  By
 construction the bridge between points (1,2) and (3,4) in the diagrams for
 ${\cal G}^{\rm wl}_{2,2}({\bf r}_0|x_1,x_2,x_3,x_4)$ can consist of a
 Green's function but not of a correlator. In panel c of Fig. \ref{95LP-01}
 the coordinates of the Green's function are denoted by $x_a,x_b$. On the
 two sides of the $(a-b)$ bridge there exist three-point objects. These
 objects have an exact resummed form in terms of the dressed vertices A and
 B which were briefly introduced in paper I. Note that we have a freedom in
 the diagrammatic representation as to whether to include the bridge itself
 together with the object on the right or on the left. In the former case
 the resulting object on the right is ${\cal G}_{1,2}({\bf
 r}_0|x_a,x_3,x_4)$ which was defined in (\ref{newG12}). In the latter
 case, which is the convention that we choose, see panel b, the resulting
 object on the left side is ${\cal G}_{2,1}({\bf r}_0|x_1,x_2,x_b)$ which
 is defined in (\ref{newG21}). In order to display the bridge $(a-b)$
 explicitly we introduce in panel c a new object denoted as ${\cal D}({\bf
 r}_0|x_1,x_2,x_a)$. This object has two "entries" 1 and 2 starting with
 propagators and one ``exit" denoted $x_a$ which ends with a vertex, panel
 d. We will use an empty small circle to denote the position of a vertex.
 This is done to distinguish a vertex from a propagator leg.

 Next, as explained in paper II, we need to identify two-eddy reducible and
 two-eddy irreducible diagrams. The first type are diagrams that can be
 split into a left and a right part by cutting two propagators. For $G_2$
 these propagators can only be Green's functions. This is the cross section
 denoted by (a) in Fig. \ref{95LP-02}. In the present case we can also have
 cross sections of type (b) with one Green's function and one propagator.
 There cannot be cross sections with two correlators because we always have
 at least one principal path of Green's functions that connects the left
 part of the diagrams to its right part. Finally we can have a cross
 section of type (c) in which two Green's functions appear in opposite
 orientations.

 In summary, the structure of the diagrammatic series can be described as
 follows:  we get ladder diagrams that have alternating Green's functions
 and double-correlators in any possible order, as long as there is at least
 one Green's function between two rungs which carries the principal path.
 The ladder diagrams can have alternating rungs of three types, depending
 on the type of propagators preceding the rung. The three types of rungs
 are sums of two-eddy irreducible diagrams that are denoted
 $\Sigma_{(1,3)}$, $\Sigma_{(2,2)}$ and $\Sigma_{(3,1)}$ respectively.
 Graphically they are represented by a dashed bar, empty bar and doubly
 dashed bar respectively.  The first one has three  straight tails and one
 wavy tail, the second  one has two straight and two wavy tails and the
 last has three wavy and one straight tail.  The series for
 $\Sigma_{(n,m)}$ are shown in Fig. \ref{95LP-03} and Fig. \ref{95LP-04}.
 Fig.2 shows the diagrams for $\Sigma_{(2,2)}$. This series is composed of
 all the diagrams in Fig.10a of paper II (and diagrams 1, 2 and 5 in Fig.
 \ref{95LP-03} are examples of those) in addition to new diagrams like
 diagrams 3 and 4 in Fig. \ref{95LP-03}. All the old diagrams have a
 horizontal principal cross section that cuts through correlators only. The
 new diagrams contain the split in the principal path, and the principal
 cross section through correlators turns up or down by $90^0$. The series
 for the two other two-eddy irreducible mass operator $\Sigma_{(1,3)}$ and
 $\Sigma_{(3,1)}$ are shown in Fig. \ref{95LP-04}. The diagrams in these
 series have no principal cross section that cuts through correlators only.
 \subsubsection{Resummation of the strongly linked contributions}
 As in the case of the ladder diagrams of $G_2$ we can resum also in
 the case of ${\cal G}_{2,2}^{\rm sl}$ all the ladder diagrams into
 integral equations. The resummed ladders are shown in Fig. \ref{95LP-05}.
 The diagrammatic notation of ${\cal G}_{2,2}^{\rm sl}$ is an empty circle
 with two wavy and two straight lines. It has three different types of
 contributions.  First come the reducible contributions that we denote by
 ${\cal G}^{(0)}_{2,2}$ and are represented by the unlinked diagrams (1)
 and (2) of Fig. \ref{95LP-02}.  The first two ladders (3) and (4) in Fig.
 \ref{95LP-02} are identical to the RHS of the resummed equation for $G_2$.
 The next term on the RHS in Fig. \ref{95LP-02} are new. The resummation of
 the ladders is shown in Figs. \ref{95LP-05}. On the RHS of the diagrammatic
 equation we find a Green's function which is defined in (\ref{newG31}) as
 ${\cal G}_{3,1}$ and whose graphic notation has a crossed circle with
 three wavy and one straight tails. This object is again resummed in terms
 of itself and ${\cal G}_{2,2}$ as shown in Fig.5b.

 In conclusion we learn that the 4-point Green's functions ${\cal
 G}_{m,n}$ satisfy equations that contains linear operators acting on
 ${\cal G}_{m,n}$ and inhomogeneous terms that are products of $G$'s and
 the weakly linked contributions. To be sure, the linear operators are
 themselves functionals of ${\cal G}_{m,n}$, so that our equations are in
 fact nonlinear. If we expand the solutions of these inhomogeneous
 nonlinear equations around the inhomogeneous terms we generate back the
 initial diagrammatic expansion. However, now we can also explore the
 ``{\it homogeneous}" solutions of these equation that are obtained by
 discarding the inhomogeneous terms. Such solutions, if they exist, are
 manifestly nonperturbative effects. We will need to show {\it a
 posteriori} that these nonperturbative solutions are much larger than the
 solutions that can be found from perturbative analysis. Indeed, in paper
 II it was shown in detail that $G_2$ has such a nonperturbative
 homogeneous solution with the following property: when the first two
 coordinates $r_1$ and $r_2$ are of the same order (say $r$) and much
 smaller than the last two coordinates $r_3$ and $r_4$ (which are of the
 order of $R$) then
  \begin{equation}
 \nabla^2_rG_2(r,r,R,R)\sim r^{-\Delta} \qquad  {\rm for} \  R \gg r \ .
 \label{diverge}
 \end{equation}
 Since the same terms appear in the equation for ${\cal G}_{2,2}$ we know
 that such a divergence (with $r\to 0$) must appear also in all ${\cal
 G}_{n,m}$.  (This of course hinges on the assumption that the new terms in
 Fig. \ref{95LP-05} do not contribute an exact cancellation; in view of
 their different nature we judge this possibility unlikely). The new terms
 may have	a stronger or a weaker divergence, and we turn now
 therefore to an exact calculation of the value of $\Delta$.
 \subsection{Calculation of $\Delta$}
 In order to evaluate $\Delta$ we establish a fundamental identity which is
 \begin{equation}
 {\delta F_{\alpha\beta}({\bf r}_0|x_1,x_2)\over \delta
 D_{\gamma\delta}(x_3,x_4)}
 ={\cal G}_{2,2}^{\alpha\beta\gamma\delta}({\bf r}_0|x_1,x_2,x_3,x_4) \ ,
 \label{funident}
 \end{equation}
 where the covariance $D$ is the correlation of the perturbations,
 \begin{equation}
 D_{\gamma\delta}(x_3,x_4)=\left<h_\gamma(x_3)h_\delta(x_4)\right> \ .
  \label{covar}
 \end{equation}
 The identity is proved most easily using the path integral formulation as
 reviewed in paper I. In terms of the functional ${\cal Z}({\bf l},{\bf
 m})$ of Eq.  (I--3.12) the second order Green's function is
 \begin{eqnarray}
 &&{\cal G}_{2,2}^{\alpha\beta\gamma\delta}({\bf r}_0|x_1,x_2,x_3,x_4)
 \nonumber\\
 &=& -\left<w_\alpha({\bf r}_0|x_1)
 w_\beta({\bf r}_0|x_2) p_\gamma(x_3) p_\delta(x_4)\right> \ . \label{G2p}
 \end{eqnarray}
 On the other hand we see from Eq. (I-3.14) that the derivative with
 respect to $D_{\alpha\beta}$ brings down $ip_\alpha p_\beta$:
 \begin{equation}
 {\delta I_0\over \delta D_{\gamma\delta}(x-y)}
 = ip_\gamma(x) p_\delta(y)\ . \label{derD}
 \end{equation}
 This means that the functional derivative of $\left<w_\alpha
 w_\beta\right>$ with respect to $D_{\gamma\delta}$ is precisely the RHS of
 (\ref{G2p}). This is the proof of the identity.

 Rewrite now the identity in the form
 \begin{eqnarray}
 &&\delta F_{\alpha\beta}({\bf r}_0|x_1,x_2)
 \nonumber\\
 &=& \int dx_3dx_4{\cal G}_{2,2}^{\alpha\beta\gamma\delta}
 ({\bf r}_0|x_1,x_2,x_3,x_4)\delta D_{\gamma\delta}(x_3,x_4) \ .
 \label{delF}
 \end{eqnarray}
 In this form this is a relation of the response $\delta F$ in the velocity
 correlator $F$ to a variation in the correlator of the random forcing
 $\delta D$.  It is clear now that if the random forcing is limited to
 scales $r_3,r_4\gg r_1,r_2$, the existence of flux equilibrium with a
 scaling solution for $F_{\alpha\beta}({\bf r}_0|x_1,x_2)$ means that the
 variation $\delta F_{\alpha\beta}({\bf r}_0|x_1,x_2)$ must be proportional
 to $F_{\alpha\beta}({\bf r}_0|x_1,x_2)$ itself. We can have a change in
 the amplitude but not in the functional form:
 \begin{eqnarray}
 \delta F_{\alpha\beta}({\bf r}_0|x_1,x_2)
  &\propto & F_{\alpha\beta}({\bf r}_0|x_1,x_2) \nonumber\\
 && {\rm for} \ r_1,r_2 \ll r_3,r_4 \ .
 \label{delFF}
 \end{eqnarray}
 To understand what are the requirements of the variation $\delta
 D_{\gamma\delta}(x_3,x_4)$ that guarantee the validity of the universal
 behaviour (\ref{delFF}) let us write again the Wyld equation (II--2.9). We
 will use economic notation, such that the vector index carries implicitly
 also the position coordinate. Repeated indices  must be summed upon and
 the convention is that this sum also requires integration over the
 intermediate position coordinates. This convention allows us to write the
 Wyld equation as
 \begin{equation}
 F_{\alpha\beta} = G_{\alpha\gamma}[D_{\gamma\delta}+\Phi_{\gamma\delta}]
 G_{\beta\delta} \ . \label{wyld}
 \end{equation}
 The variation $\delta D_{\gamma\delta}$ causes a variation in $F$. In our
 convention
 \begin{equation}
 F_{\alpha\beta}+\delta F_{\alpha\beta} = G_{\alpha\gamma}[D_{\gamma\delta}+
 \delta D_{\gamma\delta}+\Phi_{\gamma\delta}]
 G_{\beta\delta} \ . \label{delwyld}
 \end{equation}
 Next note that $\Phi_{\gamma\delta}({\bf r}_0|x_a,x_b)$ can be exactly
 expressed as a second derivative of the 4-point correlation function
 $F_4(x_a,x_a,x_b,x_b)$, cf. Eq. (I-4.5). Consequently
 \begin{equation}
 \Phi_{\gamma\delta}({\bf r}_0|x_a,x_b)\sim r_{ab}^{\zeta_4-2} \ . \label{phi}
 \end{equation}
 We know that $\zeta_4$ is expected to be considerably smaller than 2. (The
 K41 estimate is $\zeta_4=4/3$, whereas experimentally one finds
 $\zeta_4\sim 1.2$). Thus $\Phi_{\gamma\delta}({\bf r}_0|x_a,x_b)$ is {\it
 growing} when the coordinates become smaller, whereas $\delta
 D_{\gamma\delta}$ is restricted to the large scales and is {\it decaying}
for smaller coordinates. We expect therefore that $\delta F$ will be
 proportional to $F$ and the constant of proportionality is determined by
 the boundary conditions at large scales where both $D$ and $\delta D$ are
 not negligible.

 For future reference we should note at this point that the proportionality
 of $\delta F^ {\alpha\beta}$ and  $F^{\alpha\beta}$ is not restricted only
 to their scaling exponents.  In fact the two quantities have the same
 tensor structure. In other words they are the same function of ${\bf
 r}_1-{\bf r}_0$ and ${\bf r}_2-{\bf r}_0$ up to constants that are
 independent of the tensor indices.  To complete the argument choose now
 $t_1=t_2$. Next notice the fact that
 \begin{equation}
 \bbox{\nabla}_1\cdot\bbox{\nabla}_2 F_{\alpha\beta}
 ({\bf r}_0|{\bf r}_1,{\bf r}_2)\propto
 r_{12}^{\zeta_2-2} \ . \label{scaling}
 \end{equation}
 Now restrict $\delta D_{\gamma\delta}(x_3,x_4)$ to $r_3,r_4\sim R\gg
 r_1,r_2$.  Applying the operator $\bbox{\nabla}_1\cdot\bbox{\nabla}_2$ to
 (\ref{delF}) we conclude that
 \begin{equation}
 \bbox{\nabla}_1\cdot\bbox{\nabla}_2{\cal G}_{2,2}^{\alpha\beta\gamma\delta}
 ({\bf r}_0|r_1,r_2,x_3,x_4) \propto r_{12}^{\zeta_2-2} \ . \label{Gdel}
 \end{equation}
 Note that the RHS is a function of $r_{12}$ only; the reason for this is
 that under the derivatives the dependence on $r_0$ disappears. Accordingly
 the restriction of validity of this result is not necessarily $r_1,r_2\ll
 R$ but just $r_{12}\ll R$.  Comparing with (\ref{diverge}) we reach the
 central result of this section:
 \begin{equation}
 \Delta = 2-\zeta_2 \ . \label{scarel}
 \end{equation}

 Finally we can argue that the application of the operator
 $\bbox{\nabla}_1\cdot\bbox{\nabla}_2$ to ${\cal G}_{3,1}$ will give rise
 to the same exponent $\Delta$ as in (\ref{Gdel}). To see this apply the
 operator to the two equations in Fig.5. Suppose that the divergence
 associated with ${\cal G}_{3,1}$ is stronger than $\Delta$. This will
 immediately force the divergence of the LHS of Fig.5a to be stronger than
 $\Delta$, in contradiction with our exact result.  In the same manner the
 divergence of $\bbox{\nabla}_1\cdot\bbox{\nabla}_2{\cal G}_{3,1}$ cannot
 be weaker than $\Delta$ due to the equation in Fig.5b. We thus conclude
 that also
 \begin{equation}
 \bbox{\nabla}_1\cdot\bbox{\nabla}_2{\cal G}_{3,1}^{\alpha\beta\gamma\delta}
 ({\bf r}_0|r_1,r_2,x_3,x_4) \propto r_{12}^{\zeta_2-2} \ . \label{Gdel31}
 \end{equation}
 \subsection{Three Point Objects and the Weakly Linked inhomogeneous
 Contributions}

 Equation (\ref{scarel}) was derived by neglecting the inhomogeneous part
 of the equation for ${\cal G}$. In order to be sure that the inhomogeneous
 part is not important we need to evaluate now the weakly linked
 contributions and compare them with (\ref{Gdel}) and (\ref{Gdel31}). We
 are not going to perform the evaluation with the same care as we did in
 the computation of $\Delta$. All that we need is to show that these terms
 are negligibly small. We will show this by assuming that the diagrams have
 the property of rigidity which was demonstrated in paper II order by
 order. This will be sufficient since we will be able to show that there
 exists a large gap in the value of the scaling exponents of the
 homogeneous and the inhomogeneous terms. Such a large gap is not expected
 to be swamped by non-perturbative effects.

 To prepare for this evaluation we need first the diagrammatic
 representation of three-point objects. These are the 3-point correlator
 $F_3$, and the Green's functions ${\cal G}_{2,1}$ and ${\cal G}_{1,2}$.
 All the three point objects can be represented with the help of the three
 types of dressed vertices which were denoted in paper II as A, B and C
 respectively, see paper II Fig. 7.  We remind the reader that vertex A is
 a junction of one straight and two wavy lines, vertex B is the junction of
 two straight and one wavy line, whereas vertex C is the junction of three
 straight lines. The diagrammatic representation of the triple correlation
 function $F_3$ in terms of these vertices is shown in Fig. 7 of paper II,
 and is reproduced in more compact form in Fig. \ref{95LP-06}, panels b,c.
 In the same manner the Green's function ${\cal G}_{2,1}$ is represented
 here in Fig. \ref{95LP-01} panel c.

 Consider now the weakly linked diagrams for ${\cal G}_{2,2}$ which are
 resummed into the form shown in the diagram in Fig. \ref{95LP-01} panel b.
In the limit of $r_1\sim r_2\sim r\to 0$ and $r_3\sim r_4\sim R$,
 rigidity means that the integral over $r_b$ contributes mostly in the
 region $r_b\sim R$. We thus need to understand the $r$ dependence of
 ${\cal G}_{2,1}({\bf r}_0|x_1,x_2,x_b)$ in the asymptotic situation
 $r_1,r_2\ll R$. Looking at Fig. \ref{95LP-01} panels c, d we see that due
 to rigidity the vertices A and B will contribute mostly in the region
 $r_a\sim r$.  In K41 scaling the exponent of  $r$ is found by noticing
 that the correlator contributes $r^{\zeta_2}$, and the Green's function
 together with the integration over $x_a$ contribute $r^z$. The Green's
 function of the bridge $a-b$ in Fig. \ref{95LP-01} panel c contributes
 $r/R^4$, or $r^1$. The vertex A itself in diagrams 1 and 2 in Fig.
 \ref{95LP-01} panel d gives $r^{-1}$, and in total we find $r^{\zeta_2+z}$
 for these contributions.  The same arguments give the same asymptotic
 behaviour of diagram 3 shown in Fig. \ref{95LP-01} panel d. This behaviour
 should be compared with $r^{\zeta_2}$ for the homogeneous term, justifying
 the neglect of the inhomogeneous contribution.  Note that small
 corrections to the exponents are not expected to erase the large gap
 $z\simeq 2/3$.
 \section{2-Point Fusion Rule for 4-Point Correlation Functions: the
 Equation for the Exponent $\mu$}

 In this section we use the fact that we know exactly the ultraviolet
 exponent $\Delta$ to derive the fusion rule for 4-point correlations. One
 immediate consequence of these rules will be a derivation of the scaling
 law (\ref{mu}).  The $n$-point simultaneous correlation function of
 Eulerian velocity differences is defined as
 \begin{eqnarray}
 && F_n^{\alpha\beta\dots\omega}({\bf R}_1,{\bf R}_2\dots{\bf R}_n)
 \nonumber\\
 &\equiv & \left<\delta u_{\alpha}({\bf r}_0|{\bf R_1},t)
 \delta u_{\beta}({\bf r}_0|{\bf R_2},t)\dots
 \delta u_{\omega}({\bf r}_0|{\bf R_n},t)\right> \ . \label{Fn}
 \end{eqnarray}
 where $\delta {\bf u}$ was defined by (\ref{delu}) such that ${\bf
 R}_j={\bf r}_j-{\bf r}_0$. By ``fusion rules" we mean the scaling
 structure of $F_n$ when two or more of the vector separations ${\bf R}_j$
 tend to zero or to each other. In this section we discuss 2-point fusion
 rules. Without loss of generality we can denote the coalescing coordinates
 as ${\bf r}_1$ and ${\bf r}_2$, and we will consider all  distances of all
 the other coordinates $r_3 \dots r_n$ from ${\bf r}_0$ to be of the
 same order $O(R)\gg r_1,r_2$. To understand the fusion rules we will use
 diagrammatic language; at the end of the discussion one can recast the
 results into formal operator algebra of a new type without reference to
 diagrams. This operator algebra (which allows multiscaling) is an exciting
 subject that will be taken up fully in a separate publication
 \cite{95LP-e}.
 \subsection{ Classification of the Diagrams for the 4-Point Correlator}
 In this section we will analyze the 4-point correlator and the dissipation
 correlation function $K_{\epsilon\epsilon}$. In particular we derive the
 scaling relation (\ref{mu}) for the exponent that governs the power law
 decay of $K_{\epsilon\epsilon}$ according to the (\ref{Kee2}). The
 derivation begins with the examination of the 4-point correlator
 $F^{\alpha\dots\delta}_4({\bf R}_1\dots{\bf R}_4)$. This quantity is
 represented as an object with 4 wavy tails, each of which represents one
 of the velocity differences with coordinates ${\bf r}_1\dots {\bf r}_4$.
 The trivial contributions to $F_4$, which we denote as $F_4^{(0)}$, are the
 three different products of $F_2$ which graphically are the ``unlinked"
 (or ``reducible") diagrams which stem from the Gaussian decomposition.
 The next set of diagrams that contributes to the 4-point correlator is
 shown in Fig. \ref{95LP-06}.  These are all the weakly linked diagrams in
 which the coordinates $x_1,x_2$ are linked to the coordinates $x_3,x_4$
 via one propagator. As in the case of ${\cal G}_{n,m}$ we cannot have
 repeated one-propagator bridges since such contributions are resummed in
 the Dyson-Wyld line resummation. There are two possible types of bridges:
 (i) the bridge ends with a straight line and (ii) the bridge ends with a
 wavy line. In case (i) the diagrams can be resummed into diagram 1 of Fig.
 \ref{95LP-06} panel a. In case (ii) the diagrams resum into diagram 2. The
 left part of diagram 1 is the Green's function ${\cal G}_{2,1}$. The
 diagrammatic presentation of this object is shown in Fig. \ref{95LP-01}
 panels c and d. The right part of diagram 1 is the three point object
 $\cal A$ shown in panel b of Fig. \ref{95LP-06} in terms of the dressed
 vertices A,B and C, and the dressed propagators.  The left part of diagram
 2 in Fig. \ref{95LP-06} is the third order correlation function $F_3$.
 Its diagrammatic representation in terms of dressed vertices and
 propagators is given by Fig.7 of paper II, and in a more compact form in
 Fig. \ref{95LP-06} panel c. We find again in this case the same
 three-point objects $\cal A$ and $\cal D$ that appeared already in panel b
 of this figure and in panel d of Fig. \ref{95LP-01}.

 In addition to these weakly linked diagrams we have the strongly linked
 diagrams appearing in Fig. \ref{95LP-07} panel b. All these are diagrams
 whose resummed parts are connected via two-propagator bridge. The strongly
 linked diagrams are constructed as follows: locate the principal cross
 section of a diagram in the infinite expansion of $F_4$. Move from the
 principal cross section to the right and to the left until you find the
 first two-eddy reducible link, which is going to form the two-propagator
 bridge. The object between these two bridges is a contribution towards
 one of the resummed central objects on the RHS of the equation in Fig.
 \ref{95LP-07}, and see Fig. \ref{95LP-08} for the beginning of the
 diagrammatic expansion for these objects. At this point collect all the
 diagrams with the same central object and resum them to the left and to
 the right. The result of this resummation is shown in Fig. \ref{95LP-07}.
 The four point objects on the right and on the left of the central objects
 are exactly the same as those appearing in Fig. \ref{95LP-05}.
 \subsection{Asymptotics}
 \subsubsection{Strongly Linked Diagrams}
 Assume that $r_1,r_2\sim r$ are much smaller than $r_3,r_4\sim R$ but
 still in the inertial interval. Denote the structure whose outer
 coordinates are ${\bf r}_a,{\bf r}_b ,{\bf r}_3,{\bf r}_4$ as $\Psi_m({\bf
 r}_a,{\bf r}_b,{\bf r}_3,{\bf r}_4)$ with $m=1\dots 4$ according to the
 enumeration of the diagrams in Fig. \ref{95LP-07}. Diagram (1--4) can now
 be written analytically as
 \begin{eqnarray}
 (1) &=& \int dx_a dx_b{\cal G}_{2,2}({\bf r}_0|x_1,x_2,x_a,x_b)
 \Psi_1({\bf r}_a,{\bf r}_b,{\bf r}_3,{\bf r}_4)
 \ , \nonumber \\ &&
 \label{1} \\
 (2) &=& \int dx_a dx_b{\cal G}_{3,1}({\bf r}_0|x_1,x_2,x_a,x_b)
 \Psi_2({\bf r}_a,{\bf r}_b,{\bf r}_3,{\bf r}_4)
 \ ,  \nonumber \\ &&
 \label{2} \\
 (3) &=& \int dx_a dx_b{\cal G}_{2,2}({\bf r}_0|x_1,x_2,x_a,x_b)
 \Psi_3({\bf r}_a,{\bf r}_b,{\bf r}_3,{\bf r}_4)
 \ ,  \nonumber \\ &&
 \label{3} \\
 (4) &=& \int dx_a dx_b{\cal G}_{3,1}({\bf r}_0|x_1,x_2,x_a,x_b)
 \Psi_4({\bf r}_a,{\bf r}_b,{\bf r}_3,{\bf r}_4)
 \ .  \nonumber \\ &&
 \label{4}
 \end{eqnarray}
 Suppose that we can argue that all the functions $\Psi_m({\bf r}_a,{\bf
 r}_b,{\bf r}_3,{\bf r}_4)$ contribute at $r_a,r_b$
 smaller contributions than those of $\Phi({\bf r}_a,{\bf r}_b)$ in Eq.
(\ref{delwyld}).
 If so, we can think of these $\Psi$ functions as generalized perturbations
 $\delta D$ that leave the $r_{12}$  dependence universal. According to
 (\ref{delF}) and (\ref{delFF}) the main $r_{12}$ dependence comes from
 diagrams (1) and (3) and it is $r_{12}^{\zeta_2}$.

 To demonstrate that this is indeed the case we can look at a typical
 diagrammatic contribution to any of the diagrams in Fig. \ref{95LP-07}.
 For example, the first contribution to $\Psi_1({\bf r}_a,{\bf r}_b,{\bf
 r}_3,{\bf r}_4)$ is shown in Fig. \ref{95LP-08}. This particular example
 comes from diagram (1) in Fig.\ref{95LP-08}, panel b upon taking diagram
 (1) for the rung in Fig. \ref{95LP-02}.  Take $r_a\sim r_b$ and examine
 the dependence on $r_a$. The two vertices contribute $r_a^{-2}$. The
 correlator in between is worth $r_a^{\zeta_2}$.  The two stretched
 correlators connecting to larger coordinates contribute $r_a^{2\zeta_2}$.
 The two time integrals in the vertices $c$ and $d$ contribute $r_a^{2z}$
 where $z$ is the dynamic scaling exponent, $\tau(r)\propto r^z$ (and see
 details in paper I). In total we have $r_a^{\beta}$ with
 $\beta=3\zeta_2+2z-2$. This has to be compared with the exponent $\alpha$
 of $\Phi({\bf r}_a,{\bf r}_b)$ which is $\alpha=\zeta_4-2$ according to
 (\ref{phi}).  In K41 evaluation $\beta=4/3$ whereas $\alpha=-2/3$. The gap
 is so large that small corrections to K41 cannot change the fact that the
 exponent of $\Phi$ is dominant, as we want. Note that this demonstration
 is perturbative (order by order with only the lowest order done
 explicitly), but again the magnitude of the gap is sufficient to validate
 the claim.

 Exactly the same arguments hold for the contribution (\ref{3}). The
 contributions (\ref{2}) and (\ref{4}) relate to the asymptotic behaviour
 of the other Green's function ${\cal G}_{3,1}$.  It is clear however that
 due to the result in (\ref{Gdel31}) the exponents are the same.
 \subsubsection{Weakly Linked Diagrams}
 The analysis of the weakly linked diagrams in this case follows closely
 the discussion of section 3C. The diagrams can be resummed as shown in
 Fig.\ref{95LP-06}, panel b. We have two contributions; in diagram (a) we
 have on the left $F_3$ and on the right the object defined in the second
 line of the figure. In diagram (2) we have ${\cal G}_{2,1}$ (shown in
 Fig. \ref{95LP-07}) on the left and the object defined in the third line
 of Fig. \ref{95LP-06}, panel b. In the limit $r_1,r_2\sim r\ll r_3,r_4\sim
 R$ the integrals over $r_b$ in both diagrams contributes mainly in the
 region $r_b\sim R$.  Accordingly in diagram (2) we have the same
 situation that was discussed in section 3C, and see Fig.\ref{95LP-04},
 panel b.  Diagram (1) on the other hand may be bounded from above by the
 asymptotics of $F_3$ with one large and two small coordinates. Under the
 property of rigidity one can see that this correlator is independent of
 the large coordinate, and is therefore proportional to $r^1$. Again we
 have a gap, and this gap can only increase if we take into account the
 time integral of the $x_b$ vertex. Thus the weakly linked diagrams can be
 again disregarded.

 \subsection{2-Point Correlation of the Energy Dissipation}
 At this point we can employ our results to evaluate the correlation
 function (\ref{Kee}).  We first write this quantity in a way that makes
 its relation to $F_4$ clear:
 \begin{eqnarray}
 &&	K_{\epsilon\epsilon}(R) =
 	\nu^2\lim_{r_{12},r_{34}\to 0} \lim_{r_{13}\to R}
 \nabla _{1\alpha}\nabla _{2\beta}
 	\nabla_{3\gamma}\nabla_{4\delta} \nonumber \\
 &\times &\Big[F^{\alpha'\beta'\gamma'\delta'}_4
 ({\bf r}_1,{\bf r}_2,{\bf r}_3,{\bf r}_4)-
 	F^{\alpha'\beta'}({\bf r}_1,{\bf r}_2)F^
 {\gamma'\delta'}({\bf r}_3,{\bf r}_4)\Big] \nonumber \\
 &\times &\Big[\delta_{\alpha\beta}\delta_{\alpha'\beta'}
 +\delta_{\alpha\beta'} \delta_{\alpha'\beta}\Big]
 \Big [\delta_{\gamma\delta}\delta_{\gamma'\delta'} +\delta_{\gamma\delta'}
 \delta_{\gamma'\delta}\Big]\ .
 \label{limK}
 \end{eqnarray}
 This is a generalization of the situation discussed above in the sense
 that we have two pairs of coalescing points, $r_1,r_2$ and $r_3,r_4$. By
 examining Fig.8a we can see that the symmetry allows us to treat each
 coalescing pair separately. In the limit $r_{12}$ and $r_{34}$ being much
 smaller than $R$ each pair of derivatives $\nabla_1\nabla_2$ and
 $\nabla_3\nabla_4$ will contribute a divergent term proportional to
 $r_{12}^{-\Delta}$ and $r_{34}^{-\Delta}$ respectively. The dependence on
 $R$ can be found knowing that the overall exponent of $F_4$ is $\zeta_4$.
  Thus the over all scaling exponent of the quantity in the RHS
 (\ref{limK}) must be $\zeta_4-4$.  Using the fact that $\Delta=2-\zeta_2$
 we conclude that
 \begin{eqnarray}
 && \nabla _{1}\nabla _{2}\nabla _{3}\nabla _{4}F_4
 ({\bf r}_1,{\bf r}_2,{\bf r}_3,{\bf r}_4)
 \label{nice} \\
 &&\qquad\qquad \sim\bar\epsilon^{4/3}[r_{12}r_{34}]^{-\Delta}
 R^{\zeta_4-2\zeta_2} \ell^{4/3-\zeta_4}
 \ , \nonumber
 \end{eqnarray}
 where for simplicity we suppressed the vector indices and we made the
 relation dimensionally correct by introducing some renormalization scale
 $\ell$. Note that for K41 scaling exponents the factor
 $\bar\epsilon^{4/3}$ fixes the correct dimensionality of the RHS. For
 anomalous exponents one needs a renormalization scale to take
 care of the difference between the K41 and the actual value of the
 exponent $\zeta_4$. It will be shown later that this renormalization scale
 must the outer scale $L$. At this point it does not matter.  The
 divergence in the limit indicated in (\ref{limK}) should be understood in
 light of the full theory for $F_4$, in which the $\nu$-diffusive terms are
 explicit. The role of these terms is precisely to truncate the divergence
 that is implied by (\ref{limK}). As a consequence the divergence is only
 applicable in the inertial range with $r_{12},r_{34}> \eta$, whereas in
 the dissipative regime the divergence disappears.  Thus for evaluating
 $K_{\epsilon\epsilon}(R)$ via inertial range values we must replace the
 limit $r_{12},r_{34}\to 0$ by $r_{12}=r_{34}=\eta$. Thus we can write
 \begin{equation}
 	K_{\epsilon\epsilon}(R) \sim \bar\epsilon^{4/3}
 \eta^{-2\Delta} R^{\zeta_4-2\zeta_2}
 \ell^{4/3-\zeta_4}\ . \label{nicer}
 \end{equation}
 Finally, we need to evaluate the viscous scale $\eta$. This is done from
 the definition of $\bar\epsilon$ which is
 \begin{equation}
 \bar\epsilon\sim\nu\lim_{r_{12}\to\eta}\nabla_1\nabla_2 S_2(r_{12})  \ .
 \label{meaneps}
 \end{equation}
 In the same spirit we can write the structure function $S_2(R)$ as
 \begin{equation}
S_2(R) \sim \bar\epsilon^{2/3}R^{\zeta_2}
 \ell^{2/3-\zeta_2} \ , \label{S2}
 \end{equation}
 where again the renormalization scale $\ell$ fixes the dimensions. From
 the last two equations one can compute
 \begin{equation}
 \bar\epsilon^{1/3} \sim \nu\eta^{-\Delta}\ell^{2/3-\zeta_2} \ . \label{eta}
 \end{equation}
 Substituting this in (\ref{nicer}) we get finally
 \begin{equation}
 K_{\epsilon\epsilon}(R) \sim \bar\epsilon^2
 \left({\ell\over R}\right)^{2\zeta_2-\zeta_4}
 \ , \label{nicest}
 \end{equation}
 which is the scaling law (\ref{Kee2}) that we wanted to derive. It will
 turn out that the renormalization scale $\ell$ is the outer scale of
 turbulence $L$. From this we get Eq. (\ref{mu})
  \begin{equation}
 \mu = 2\zeta_2-\zeta_4 \ .
 \end{equation}
 \section{2-Point Fusion Rule for many-Point Correlation Functions:
 the Derivation of {$J_n$}}

 In this section we evaluate the scaling exponent of the dissipative terms
 $J_n$ (\ref{J2mfinal}) of the balance equations. The strategy is to expose
 the divergence with the ultraviolet cutoff $\eta$ for which we have an
 exact evaluation of the exponent $\Delta$. Together with the overall
 scaling which is determined by $\zeta_n$ we will be able to compute the
 $R$ dependence of $J_n(R)$. Having that we will use an argument due to
 Kraichnan to find the coefficient of the power law. We believe that our
 result for $J_n(R)$ is exact, even though some of steps taken are not
 rigorous.  Afterwards we will evaluate the interaction term $D_n(R)$ and
 use the balance equation as a non-perturbative constraint to deduce the
 scaling exponents $\zeta_n$.
 \subsection{Classification of the Diagrams as  Weakly Linked
 or Strongly Linked}
 The $n$-point correlation function $F_n$ can be represented symbolically
 as an object with $n$ wavy tails, each one representing one of the $n$
 coordinates ${\bf r}_j$ associated with a velocity difference $\delta
 u({\bf r}_0|{\bf R}_j)$. We remind the reader that for simultaneous
 correlation functions one can use either the Eulerian differences
 (\ref{delu}) or the BL-velocity differences (\ref{delw}), since they are
 the same at $t=0$. In the diagrammatic expansion for the simultaneous
 $F_n$ there appear time dependent correlations, and there the theory calls
 for the use of BL-velocity differences.

 Having two special coordinates ${\bf r}_a$ and ${\bf r}_b$ we can ask how
 the part of the diagram containing these coordinates is linked to the
 rest of the diagram. This part can be connected via one propagator, see
 Fig.   \ref{95LP-10} panel a, via two propagators, see Fig.
   \ref{95LP-10} panel b, or via three or more. As in the case of the
 4-point correlator and 4-point Green's function a one-propagator bridge
 cannot appear again between the legs carrying the designation $a, b$ and
 the body of the diagram. In total one has $n(n-1)/2$ weakly-linked
 contributions in each of which the role of the weakly linked pair is
 played by one of the available pairs of legs. In addition one has also
 double weakly-linked contributions with two bridges made of a single
 propagator which connects different pairs, etc, and see diagram 3 in Fig.
 \ref{95LP-10} panel a. Such diagrams do not play an important role in our
 analysis.  In displaying the diagrams in Fig.10a we have included the
 bridge (which is a correlator or a Green's function) in the object on the
 left of the bridge.  For this reason the object on the right of the
 bridge begins with a vertex $x_c$. According to our convention this
 vertex is denoted by a small empty circle.  The diagrams appearing in
 Fig.  \ref{95LP-10} panel a should already look familiar. In diagram 1 we
 again have a $F_3$ correlator on the left, integrated over $x_c$ with an
 $(n-1)$-point object on the right.  This is the generalization of the
 diagram 2 in Fig.  \ref{95LP-06} panel a.  In diagram 2 we have the
 3-point Green's function ${\cal G}_{2,1}$ of Eq.(\ref{newG21}) on the
 left, again integrated over $x_c$ with a different $(n-1)$-point object.
 This can be compared with diagram 1 in Fig.  \ref{95LP-06} panel a. For
 our considerations the precise nature of the object on the right is
 irrelevant.

 Similarly, all  the contributions with two propagators serving as links
 can be resummed into the objects shown in Fig. \ref{95LP-10} panel b.
 Again we included the two propagators that form the bridges in the objects
 appearing to the left of the bridges. Again this results in the $n$-point
 objects on the right being attached to the left by two vertices $x_a$ and
 $x_b$.  Their interpretation is as follows:  diagram 1 is a fourth order
 correlator linked via integrals over $x_a$ and $x_b$ to the rest of the
 diagram. In diagram 2 the linked object is ${\cal G}_{3,1}$,
 (\ref{newG31}). In diagram 3 the linked object is precisely our familiar
 second order Green's function ${\cal G}_{2,2}$.

 Links with three or more propagators have been already taken into account
 in this presentation. This classification of the diagrams is based
 on starting with the two special coordinates $x_1,x_2$ on the left, and
then moving to the right and stopping when the first one-propagator
 bridge appears. All these diagrams are resummed exactly into one of the
 contributions in Fig.10a. If there is no one propagator bridge, we start
 again from the left and monitor all the two-propagator bridges, ending with
 the last pair. All such diagrams are resummed into one of the
 contributions in Fig. \ref{95LP-10}, panel b.

 \subsection{Asymptotic Tensor   Structure and Fusion Rules}

 In this section we find the tensor structure of $F_n^{\alpha \beta \dots
 \omega}$ when the two coordinates $r_1,\ r_2$ are much smaller than the
 rest.  When all the coordinates $r_3,\dots r_n$ are of the order of $R\gg
 r$, whereas $r_1,r_2$ are small, the property of rigidity that was
 demonstrated in paper II requires that the main contribution to the
 integrations over $r_a$ and $r_b$ in the diagrams of Fig. \ref{95LP-10}
 panel b come from the region $r_a\sim r_b \sim R$. In this case we have a
 very similar situation to the one discussed before in the context of Fig.
 \ref{95LP-07} panel b. Accordingly the three objects on the left of the
 double bridge of the diagrams in Fig. \ref{95LP-10} panel b are familiar
 and have the same scaling with respect to $r$, i.e.  $r^{\zeta_2}$. The
 weakly linked contributions shown in Fig. \ref{95LP-10} panel a have the
 same objects on the left of the bridge as those shown in Fig.
 \ref{95LP-06} panel a for $F_4$, and they are irrelevant for the same
 reasons.

 Limiting our attention to strongly-linked diagrams with two propagator
 links we examine now the tensor structure of
 $F_n^{\alpha\beta\dots\omega}$. We need to keep in mind equation
 (\ref{delFF}) which means that when $r_1,r_2\to 0$  $\delta
 F^{\alpha\beta}\propto F^{\alpha\beta}$.  In light of  Fig. \ref{95LP-05}
 the objects ${\cal G}^{\alpha\beta\gamma\delta}_{2,2}$ and ${\cal
 G}^{\alpha\beta\gamma\delta}_{3,1}$ have the same asymptotics. We have
 found before that they are proportional to $F^{\alpha\beta}({\bf r}_1,{\bf
 r}_2)$.  Diagram 1 is again proportional to the same quantity in light of
 Fig. \ref{95LP-07} panel b.  We can therefore write the following fusion
 rule:
 \begin{eqnarray}
 &&\lim_{r_1,r_2 \to 0} F_n^{\alpha\beta\gamma
 \dots\omega}({\bf r}_1,{\bf r}_2,{\bf R}_3\dots {\bf R}_n)
 \nonumber \\
 &=& F^{\alpha\beta}(({\bf r}_1,{\bf r}_2) \Psi_{n-2}^{\gamma\dots
 \omega}({\bf R}_3\dots {\bf R}_n) \ .
 \label{fuse2}
 \end{eqnarray}
 Here $\Psi_{n-2}$ is a homogeneous function of its arguments when they are
 all in the inertial range. The scaling exponent of $\Psi_{n-2}$ is
 $\zeta_n-\zeta_2$. The reason for that is clear: the scaling exponent of
 $F_n$ is $\zeta_n$, but one $\zeta_2$ is already carried by
 $F^{\alpha\beta}$. The tensor structure of $\Psi_{n-2}^{\gamma\dots
 \omega}$ is not known in detail at this point, except that it conforms
 with incompressibility and isotropy. In other words, this quantity is
 independent of the vectors ${\bf r}_1,{\bf r}_2$.
 \subsection{The Evaluation of $J_n(R)$}

 At this point we are ready to evaluate the dissipative terms in the
 balance equation.  The equation to consider is (\ref{J2mfinal}). As was
 done in the context of the evaluation of $K_{\epsilon\epsilon}$ we
 evaluate the quantity in the inertial interval, and then replace the
 limits $d_\beta \to 0$ with the ultraviolet cutoff $d_\beta =\eta$. In the
 inertial interval we write
 \begin{eqnarray}
 T^{\alpha\alpha}_{2m}(d_\beta{\bf e}_\beta,d_\beta{\bf e}_ beta,{\bf R})
 &=&
 F^{\alpha\alpha} (d_\beta {\bf e}_\beta,d_\beta {\bf e}_\beta)
 \Psi^{(1)}_{2m-2}(R)  \ ,
 \nonumber \\ &&  \label{Talal} \\
 T^{\alpha\gamma\alpha\gamma}_{2m}(d_\beta {\bf e}_ \beta,d_\beta {\bf
 e}_\beta,{\bf R}) &=& F^{\alpha\gamma} (d_\beta {\bf e}_\beta,d_\beta {\bf
 e}_\beta) \Psi^{(2) \alpha\gamma}_{2m-2}({\bf R})  \,,
 \nonumber \\ &&  \label{Talgam}
 \end{eqnarray}
 where the fusion rule (\ref{fuse2}) has been used. The two functions
 $\Psi$ satisfy
 \begin{eqnarray}
 \Psi^{(1)}_{2m-2}(R) &=& A_1 R^{\zeta_{2m}-\zeta_2} \ ,
 \label{psi1} \\
 \Psi^{(2) \alpha\gamma}_{2m-2}({\bf R})
 &=& A_2 R^{\zeta_{2m}-\zeta_2} \left[\delta_{\alpha\gamma}
 -a_m {R_\alpha R_\gamma \over R^2}\right] \ ,
 \label{psi2}
 \end{eqnarray}
 with $A_1$ and $A_2$ being constants.  The reason for these forms stems
 again from the fusion rule. $\Psi^{(1)}$ and $\Psi^{(2) \alpha \beta}$ are
 scalar and 2-tensor respectively, and we wrote their general forms for
 isotropic conditions. Incompressibility dictates the value of $a_m$ but we
 do not need to compute it for our purposes. Lastly we  note that
 \begin{equation}
 F^{\alpha\gamma}(d_\beta {\bf e}_\beta,d_\beta {\bf e}_\beta) = S_2
 ^{\alpha\gamma}(d_\beta) \propto d_\beta^{\zeta_2} \left[\delta_{\alpha\gamma}
 -a_1 {R_\alpha R_\gamma \over R^2}\right] \ . \label{FS}
 \end{equation}
 Incompressibility requires the relation $a_1 = \zeta_2/(1+\zeta_2)$.

 Presently we can substitute all this knowledge into Eq. (\ref{J2mfinal}).
 One should note that in the inertial range the velocity field is not
 smooth, ($\zeta_2 < 2$) and we may run into the danger that the quantity
 computed depends on the angle between the vectors ${\bf d},{\bf R}$.
 However the procedure implied requires taking the limit such that all
 $d_\beta = \eta$. This can be checked explicitly by introducing the tensor
 structure to all the quantities appearing in the limit that indeed the
 condition that all the  components $d_\beta$  are the same guarantees that
 the limit is independent of the angle. The result of the substitution is
 \begin{equation}
 J_{2m}(R) = 2m C_{2m} \nu \eta^{-\Delta}
 R^{\zeta_{2m}-\zeta_2} \ell^{2m/3-\zeta_{2m}} \ .
 \label{J2mres}
 \end{equation}
 We again use a renormalization scale $\ell$ to fix the
 dimensions. It will turn out that this renormalization scale is the outer
 scale of turbulence $L$.

 We will see that this form of $J_n$ is sufficient for the calculation of
 the exponents $\zeta_n$ through the use of the balance equation only if
 the dependence of $\zeta_n$ on $n$ is linear (i.e in the
 $\beta$-model). If the dependence is nonlinear (multiscaling) we need to
 determine the coefficients $C_{2m}$ exactly. To do so we rewrite now Eq.
 (\ref{J2mres}) in terms of the structure functions $S_n$. This way of
 writing is compelling only when the scaling exponents are a nonlinear
 function of $n$, as will be clear in a moment.  We write $S_n$ in the
 form
 \begin{equation}
 S_{2m}(R) \sim \bar\epsilon^{3m/3} R^{\zeta_{2m}}
 \ell^{2m/3-\zeta_{2m}} \ . \label{S2m}
 \end{equation}
 This form is the generalization of (\ref{S2}) and is the most general form
 that conforms with scaling and is dimensionally correct. Using Eqs.
 (\ref{S2}) and (\ref{eta}) in (\ref{S2m}) we find the convenient
 representation
 \begin{equation}
 J_{2m}(R) = m {C_{2m} \over C_2} J_2 {S_{2m}(R)\over S_2(R)} \ .
 \label{J2malfi}
 \end{equation}
 We stress that this result is valid only when $R \gg \eta$, since we used
 the asymptotics, and only when $\zeta_n$ is nonlinear in $n$. If the
 scaling exponents are linear in $n$ we can have other contributions like
 $S_{2m+1}/S_3$ or any other ratio whose scaling exponent is $\zeta_{2m}
 -\zeta_2$. In particular (\ref{J2malfi}) is not applicable to Burgers
 turbulence.

 Finally, we will employ an idea that is due to R.H. Kraichnan \cite{94Kra}
 to argue that in the multiscaling case the coefficient $C_{2m}$ is
 $m$-independent. Begin with Eq. (\ref{J2m}) which is rewritten as an
 integral over the distribution function $P({\bf w})$:
 \begin{equation}
 J_{2m}(R) = \int d{\bf w} P({\bf w}) w^{2(m-1)}
 w_{\alpha}\left<[\nabla_r^2+\nabla_{r_0}^2] w_\alpha |{\bf w}\right> \ .
 \label{mumbo}
 \end{equation}
 Here $\left<[\nabla_r^2+\nabla_{r_0}^2]w_\alpha |{\bf w} \right>$ is the
 conditional average of $[\nabla_r^2+\nabla_{r_0}^2] w_\alpha$ conditioned
 on a given value of  ${\bf w}({\bf r}_0|{\bf r}_0+ {\bf R},t)$. The point
 to observe now is that the only way to recover our result (\ref{J2malfi})
 when $\zeta_n$ is a nonlinear function of $n$ is to demand that the
 conditional average satisfies
 \begin{equation}
 \Big<[\nabla_r^2+\nabla_{r_0}^2] w_\alpha |{\bf w}\Big> = C
 {w_\alpha({\bf r}_0|{\bf r}_0+ {\bf R},t)\over S_2(R)} \ ,
 \label{jumbo}
 \end{equation}
 where $C$ is some coefficient which is evidently independent of $m$. It
 follows that $C_{2m}$ is independent of $m$.

 Note that this result for the conditional average is only valid in the
 inertial range of scales, since it has been derived using Eq.
 (\ref{J2malfi}) which is only valid there.  Notwithstanding we can employ
 (\ref{jumbo}) right away to compute the vector quantity
 $J^\alpha_{2m+1}(R)$. Writing (\ref{Jalph}) again as an integral over the
 distribution function $P({\bf w})$ and substituting (\ref{jumbo}) we find
 \begin{equation}
 J^\alpha_{2m+1}(R) = {(2m+1)\over 2} J_2
 {S^\alpha_{2m+1}(R)\over S_2(R)} \ .
 \label{J2malfivec}
 \end{equation}
 We can thus summarize this section with a result that is valid for both
 odd and even $n$ by using the scalar counterpart of the vector quantities:
 \begin{equation}
 J_n(R) = J_2 {n S_n(R)\over 2 S_2(R)} \ . \label{Jnfinal}
 \end{equation}
 This is the final result of this section. We note that such a scaling
 formula for $J_{2m}$ was suggested by Kraichnan in the context of passive
 scalar advection (\cite{94Kra}), and was derived in \cite{95FGLP}.

 \subsection{The Dynamical Exponent $z_n$}
 As a windfall profit from the calculation of $J_n$ we can show now that the
 dynamical exponents associated with the $n$-point correlations are all the
 same. The dynamical exponent $z_n$ is defined by the assumption of scale
 invariance of the $n$-point, $n$-time correlation function of the
 BL-velocity differences in the form
 \begin{eqnarray}
 && F_n(\lambda {\bf R}_1, \lambda^{z_n}t_1,\dots \lambda {\bf R}_n ,
 \lambda^{z_n}t_n)
 \nonumber\\
 &=& \lambda^{\zeta_n} F_n( {\bf R}_1,t_1,\dots  {\bf R}_n ,t_n) \ .
 \label{zn}
 \end{eqnarray}
 The meaning of this form is that the typical time scale associated with
 $n$-point quantities scales like $R^{z_n}$ when $R$ is changed. In
 particular we can examine Eq. (\ref{balancesc}) and realize that $J_n$,
 which according to (\ref{Jnfinal}) is $nJ_2S_n(R)/S_2(R)$, gives us the
 desired time scale. In fact, the $R$ dependence of the time scale is
 determined entirely by $S_2(R)$, and therefore
 \begin{equation}
 z_n = \zeta_2	\quad {\rm for~~all~~} n \ . \label{dynamic}
 \end{equation}
 This is a non-trivial prediction that to our knowledge has never been
 tested either in experiments or simulations. It appears to be exact. It is
 interesting to notice that independently of the question of multiscaling in
 the spatial scale, the temporal scaling is simple.

 It is amusing to try to understand (\ref{dynamic}) intuitively. In doing
 so we want to separately understand why $z_2=\zeta_2$ and then why all
 $z_n$ are the same. The first finding seems to contradict the naive
 dimensional evaluation of $\tau_2(R)$ as $R/\sqrt{S_2(R)}$, which is
 the ``turn-over" time of $R$-eddies with characteristic velocity
 $\sqrt{S_2(R)}$.  This evaluation would lead to $z_2=1-\zeta_2/2$ which is
 wrong.

 Another way of thinking that leads to the right result is to estimate the
 rate of energy dissipation as the ratio of energy of $R$-motions, which is
 $S_2(R)$, by the time scale $\tau_2(R)$. Since the rate of energy
 dissipation is $R$-independent (being $\bar\epsilon$), this fixes
 $\tau_2(R)$ to scale as $R^{\zeta_2}$.

 The $n$-independence of $z_n$ is more subtle, and we postopone its
 discussion to the forthcoming paper \cite{95LP-e}. Here we just want to
 point out that this result entails a prediction about the measurement of
 dimensionless ratios of structure functions, like $S_3/S_2^{2/3}$,
 $S_4/S_2^2$ etc. in decaying turbulence. The prediction is that such
 relatios are $R$-dependent but not time dependent. We believe that this is
 not in contradiction with what is known about decaying turbulence behind a
 grid.

 \section{The Interaction Term in the Balance Equation}
 In this section we present the analysis of the interaction term $D_n$, and
 see (\ref{D2mfinal}) and (\ref{D2vecfinal}). The question that was left at
 the end of sec 2C 1 is whether the integral over ${\bf r}_1$ converges.
 Order by order analysis of the type presented in paper I indicates that
 the answer is yes. However we need now to consider the nonperturbative
 answer using what we have learned so far.

 \subsection{Locality of the integral in the interaction term}
 In order to do this we need further asymptotic properties of the functions
 $T_n$ which appear in the integral. For brevity we will suppress
 the tensor indices of these objects, and to consider even and odd $n$ in
 the same way.  The convergence of the integrals depend on $T_n({\bf
 R}+{\bf r}_1,{\bf R}+{\bf r}_1,{\bf R})$ and $T_n({\bf r}_1,{\bf r}_1,{\bf
 R})$ when $r_1 \gg R$ and when $r_1 \ll R$. So far we have only analyzed
 $T_n({\bf r}_1,{\bf r}_1,{\bf R})$ when $r_1 \ll R$. The full analysis of
 the two unknown asymptotics is as involved as the one presented above, and
 we will present them in a separate publication \cite{95LP-e}. Here
 we will simply employ the results that we need for the present
 analysis.

 Consider first $T_n({\bf R}+{\bf r}_1,{\bf R}+{\bf r}_1,{\bf R})$ for
 $r_1$ small. The analysis in \cite{95LP-e} shows that for $r_1 \ll R$
 \begin{equation}
 T_n({\bf R}+{\bf r}_1,{\bf R}+{\bf r}_1,{\bf R}) - S_n(R)
 \propto S_2(r_1) \propto r_1^{\zeta_2}
 \ . \label{Tasy}
 \end{equation}
 Next consider $T_n({\bf r}_1,{\bf r}_1,{\bf R})$ in the limit $r_1 \gg R$.
 The analysis in \cite{95LP-e} leads to
 \begin{equation}
 T_n({\bf r}_1,{\bf r}_1,{\bf R}) \propto R^{\zeta_{n-2}}
 r_1^{\zeta_n-\zeta_{n-2}}
 \ . \label{Tasyir}
 \end{equation}
 These results can be employed now in the integral for $D_{2m}$. In this
 integral we have the projection operator $\OP$, which has a delta function
 and a longitudinal part. It was demonstrated in section 2 that the delta
 function leads to the expression (\ref{D2mnop}), as if there were no
 pressure. The longitudinal part of $\OP({\bf r}_1)$ is proportional to
 $1/r_1^3$. The integral $\int d{\bf r}_1 \OP({\bf r}_1)$ by itself
 diverges logarithmically. The rest of the integrand (i.e. $\partial T_n /
 \partial r_{1\gamma}$) behaves like $r_1^{-\Delta} r_{1\gamma}$. Simple
 power counting indicates that the integral diverges on the whole in the
 ultraviolet region. In fact, this power counting is misleading, since the
 integration over the angles vanishes. The projection operator is an even
 function under the inversion of ${\bf r}_1$, whereas the leading term of
 the rest of the integrand is odd. The next term in the expansion of
 $\partial T_n / \partial r_{1\gamma}$ is even under the inversion of ${\bf
 r}_1$, and is of the order of $r_1^{\zeta_2}$. The resulting integral
 $\int dr_1 r_1^{\zeta_2-1}$ converges in the ultraviolet.

 Note that this analysis indicates that each of the two terms in the
 integral for $D_n$ converge in the ultraviolet independently. In fact, we
 see from Eqs. (\ref{Tasy}) and (\ref{Talal}) that the two terms have
 precisely the same asymptotics, and they may exactly cancel in the limit.
 In addition we see from (\ref{D2mfinal}) that the leading asymptotics of
 the two terms cancels exactly also in the infrared. These facts are
 important; we will argue below that the leading scaling behaviour for
 $D_n$ which is naively calculated from each term separately may cancel,
 and the actual scaling is determined by the next order contribution. This
 will be a mechanism for multiscaling.

 Notwithstanding the exact cancellation of the leading infrared behavior
 we need to examine the infrared convergence of the integral. Each one of
 the terms in the integrand of $D_{2m}$ has the asymptotic form $\int
 (dr_1/r_1) \partial r_1^{\zeta_n- \zeta_{n-2}}/\partial r_1$ which
 converges separately in the infrared. The difference should converge even
 faster.

 In summary, we argued here that the proof of locality of the integral for
 $D_{2m}$ extends beyond order by order considerations. Similar arguments
 allow reaching the same conclusion for the integrals in $D_{2m+1}$.
 \subsection{Rough Evaluation of $D_n$}
 The conclusion of the last subsection is that the main contribution to the
 integrals appearing in $D_n$ comes from the region $r_1\sim R$.
 Accordingly the integral can be evaluated as discussed in \cite{95LP}:
 \begin{equation}
 D_n(R) \sim {dS_{n+1}(R)\over dR} \ . \label{eval1}
 \end{equation}
 This is exactly of the form computed in Eq. (\ref{D2mnop}) for the case
 without pressure.  We will see next that if this evaluation is to be
 trusted, its unavoidable consequence is that the scaling exponents
 $\zeta_n$ are a linear function of $n$. We will see that this allows only
 K41 scaling or $\beta$-model type scaling \cite{78FSN}. It is possible
 however that (\ref{eval1}) is an overestimate; we argued above that the
 interaction term as shown in Eqs. (\ref{D2mfinal}) and (\ref{D2vecfinal})
 may have a cancellation of the leading scaling behavior which is valid for
 every one of the terms in the integrals separately. We will therefore also
 study now the next order term that will be the proper evaluation of $D_n$
 {\it if} the leading evaluation indeed cancels. We will see that the
 resulting evaluation culminates in multiscaling in close agreement with
 experimental observations. Moreover, we will argue that as far as we can
 see within the new approach developed in this series of papers, the
 scenario that involves the cancellation of the leading scaling behaviour
 of $D_n$ is the {\it only} scenario that allows multiscaling in
 turbulence. It is possible that there exists a symmetry or a sum rule that
 leads to such a cancellation but we do not have at the present time a
 theoretical proof or even a good argument as to why and how it happens.

 To evaluate the next order scaling contribution of $D_n$ we need to return
 to the diagrammatic expansion of $F_n$, Fig. \ref{95LP-10}. In the
 discussion in section 5 we explained that in the asymptotic regime of two
 small coordinates the weakly linked diagrams are negligible compared with
 the two-propagator bridged contributions, even in K41 scaling. Now however
 we are interested in these diagrams when all the coordinates are of the
 same order, and it is evident that K41-wise they all have the same scaling
 with $R$. In fact, the unlinked contributions to $F_n$ which are obtained
 from the Gaussian decomposition (i.e. all the contributions $F_p F_q$
 with $p+q=n$) also have the same K41 evaluation. In addition we have a set
 of  weakly linked contributions such as the ones displayed in Fig.
 \ref{95LP-10}, panel a. Again they have the same scaling in K41.
 Accordingly we need to think which contributions are dominant when the
 leading scaling $(R/L)^{\zeta_n}$ cancels in the evaluation of $D_n$.
 (Since we are going to show that anomalous scaling requires the
 normalization scale to be $L$, we assume this in the present discussion
 without further ado).

 The estimate of the scaling exponents of all these various contributions
 is facilitated by the fact that we are interested now in the ``local"
 situation when all the coordinates are of the order of $R$. Thus for
 example the diagrams 2  in Fig. \ref{95LP-10} panel a are redrawn in Fig.
 \ref{95LP-11}. The objects in the left grey ellipse in diagrams 1 and 2
 are representation of $F_3$ shown in panel c of \ref{95LP-06}. The object
 in the right ellipse of diagram 2 belongs to $F_{n-1}$. One can see this
 by taking $n=3$ and looking back to panel c of Fig. \ref{95LP-06}. However
 in diagram 2  we counted the $F_2$ bridge twice. The overall scaling
 exponent is therefore $\zeta_3+\zeta_{n-1}-\zeta_2$.  Using Fig.
 \ref{95LP-06} panel b it can be also seen that the diagram 1 in Fig.
 \ref{95LP-11} has the same scaling exponent. Similarly we can analyze the
 diagram 1 in Fig. \ref{95LP-10} panel a and argue that its scaling
 exponent is $\zeta_3+\zeta_{n-1}-\zeta_2$, etc.

 Now we need to understand which of these contributions will take the
 lead if the main scaling $(R/L)^{\zeta_n}$ cancels. To guide our thinking
 we will assume that the scaling exponents are neither K41 nor
 $\beta$-model, but are nonlinear functions of $n$. H{\"o}lder inequalities
 then require that the increments between $\zeta_n$ and $\zeta_{n-1}$ will
 be nonincreasing functions of $n$, i.e.
  \begin{equation}
 \zeta_{n+1} -
 \zeta_n \leq \zeta_n-\zeta_{n-1} \ .
 \label{holder}
 \end{equation}
 With these inequalities one sees that the unlinked contributions which
 scale with $(R/L)^{\zeta_p+\zeta_q}$ are always smaller than the weakly
 linked contributions, and that of all the weakly linked contributions the
 leading one is the one which we singled out in Fig. \ref{95LP-10} panel a,
 with the scaling exponent $\zeta_3+\zeta_{n-1}-\zeta_2$.

 The meaning of this result is that instead of evaluating $T_n$ in the
 integrals for $D_n$ as $S_n(R)$ we need to evaluate it as $S_{n-1}(R)
 S_3(R)/S_2(R)$. Correspondingly the evaluation (\ref{eval1}) changes to
 \begin{equation}
 D_n(R) = d_n(\zeta_n) {S_n(R)S_3(R)\over R S_2(R)} \ , \label{eval2}
 \end{equation}
 where $d_n(\zeta_n)$ is a coefficent which depends on the numerical value
 of the scaling exponent.
 \section{Analysis of the Balance Equation as a Nonperturbative Constraint}
 At this happy moment we can use all the knowledge accumulated so far to go
 back to the balance equations (\ref{balsc}) and (\ref{balvec}) that we
 rewrite in the form
 \begin{equation}
 D_n(R) =J_n(R) \ , \label{bal}
 \end{equation}
 where the forcing term is not displayed because it is negligible, cf.
 section 2C3.  The evaluation of $J_n(R)$ is given in (\ref{Jnfinal}). The
 evaluation of $D_n$ is either (\ref{eval1}) or (\ref{eval2}), depending
 whether there is oris not cancellation in the leading scaling
 behaviur of $D_n$. We will show now that option (\ref{eval1}) leads
 inevitably to linear scaling, whereas option (\ref{eval2}) leads to
 multiscaling.
 \subsection{Scenario of Linear Scaling: Burgers Turbulence and the
 $\beta$-model}
 The evaluation (\ref{eval1}) is exactly correct only when there is no
 pressure term and the projection operator is a delta function. This is the
 situation for example in Burgers turbulence \cite{94Got}. It may or may
 not be a proper evaluation of $D_n$ also in the case of Navier-Stokes
 turbulence, as discussed above. We examine now the consequences of this
 evaluation when substituted in the balance equation.  Substituting
 (\ref{eval1}) and (\ref{Jnfinal}) in (\ref{bal}) we find
 \begin{equation}
 {S_{n+1}(R)\over R} \sim \bar\epsilon{ S_n(R)\over S_2(R)} \ . \label{linsc1}
 \end{equation}
 For $n=2$ we recover the known result that $S_3(R)\sim\bar\epsilon R$.
 Accordingly we can rewrite (\ref{linsc1}) as
 	 \begin{equation}
 S_{n+1}(R)S_2(R) \sim  S_n(R)S_3(R) \ . \label{linsc2}
 \end{equation}
 In terms of the scaling exponents this results reads
 \begin{equation}
 \zeta_{n+1}+\zeta_2 \sim  \zeta_n+\zeta_3  \ . \label{linexp}
 \end{equation}
 The only solution of this equation is the linear law $\zeta_n=a+bn$, where
 $a$ and $b$ are some constants. Knowing that $\zeta_3=1$ and using our
 scaling law (\ref{mu}) which is $\mu=2\zeta_2-\zeta_4$ we find that the
 only solution is
 \begin{equation}
 \zeta_n={n\over 3}-\mu{(n-3)\over 3} \ . \label{beta}
 \end{equation}
 It is interesting to note that this result is identical to the prediction
 of the $\beta$-model \cite{78FSN}, coefficients and all. This should not
 surprise us too much. After all, once we have a linear dependence two
 constraints fix the linear law completely. Note that this law includes as
 a special case the exponents of Burgers turbulence which are $\zeta_n=1$
 for all $n$ \cite{94Got}. This is obtained from (\ref{beta}) when $\mu=1$.
 Nevertheless the full analysis of the Burgers equation using the
 techniques developed in this series of papers needs special attention due
 to the importance of the incompressibility constraint in so many of our
 calculations. The Burgers case deviates so strongly from K41 that the
 issues of locality and rigidity of the various diagrams needs to be
 assessed separately \cite{95Pol}.

 We can also show now that (\ref{beta}) implies that the renormalization
 scale is the outer scale of turbulence $L$ as claimed before. Eq.
 (\ref{beta}) was derived by asserting that the Gaussian contribution to
 $J_n$ is negligible compared to the connected ladder contributions which
 led to (\ref{Jnfinal}). The Gaussian contributions are dominated by
 $\bar\epsilon S_{n-2}$ whose scaling exponent is $\zeta_2$. If we used
 this contribution as the leading one in the balance equation (\ref{bal})
 we would have obtained the scaling relation $\zeta_{n+1}=\zeta_3+\zeta_{n-2}$
with
 the obvious boundary condition $\zeta_0=0$.  The solution of this
 recursion relation is K41 scaling with $\zeta_n=n/3$. If this is to be
 rejected in favor of (\ref{beta}) the Gaussian contributions must be
 indeed smaller than the ones we kept. The conclusion is that
 \begin{equation}
 S_2(R)S_{n-2}(R) < S_n(R) \ . \label{ineq}
 \end{equation}
 In turn this inequality implies that
 \begin{equation}
 \left({R\over \ell}\right)^{\zeta_{n-2}} < \left({R\over
 \ell}\right)^{\zeta_n-\zeta_2}
 \ . \label{ineq2}
 \end{equation}
 Substituting the scaling exponents from (\ref{beta}) we conclude that
 $(R/\ell)^\mu < 1$, which can only happen if $R < \ell$ for any $R$ in the
 inertial inteval. This identifies $\ell$ with $L$.
 \subsection{Scenario of Multiscaling}
 Substituting (\ref{eval2}) and (\ref{Jnfinal}) in (\ref{bal}) we find
 \begin{equation}
 d_n(\zeta_n) {S_n(R)S_3(R)\over RS_2(R)} = {n J_2 S_n(R)\over 2 S_2(R)}
 = { n S_n(R)S_3(R)\over 2 RS_2(R)} \ , \label{mult1}
 \end{equation}
 where the last form is obtained from $J_2 = S_3/R$.  From the point of
 view of scaling exponents this equation is an identity.  The only way to
 compute the exponents $\zeta_n$ now is from the {\it coefficients} in the
 balance equation:
 \begin{equation}
 2d_n(\zeta_n) = n \ . \label{coefbal}
 \end{equation}
 To this aim the rough evaluation of $D_n$ in section 6B is not sufficient;
 we need to be much more precise in order to compute the $\zeta_n$
 dependence of $d_n$.

 Clearly, the computation of coefficients is exceedingly hard. Only when we
 have exact form for the functional dependence of the many point functions
 we can hope to compute the coefficient. We did have an exact form for $J_n$
 because we understood how to resum its diagrams; consequently we believe
 that we have computed the coefficient of $J_n$ exactly. $D_n$ is a different
 matter; at present we do not have an exact functional form for it. Order by
 order considerations are not helpful for this issue, and we still do not
 know how to exactly resum the diagrammatics for $D_n$. We will therefore
 try to guess the $\zeta_n$ dependence of the coefficient of $D_n$.
 \subsubsection{The Eddy-Viscosity Approximation of $D_n(R)$}
 To guide our thinking we recall some results from the theory of passive
 scalar advection \cite{94Kra,95FGLP,94LPF}.  In that problem $D_n$ had the
 form of an eddy-diffusivity operator:
 \begin{equation}
 D_n^{\rm passive}(R) = {1\over R^2}{d\over dR}R^2 h(R){d\over dR}S_n(R) \ ,
  \label{passD}
 \end{equation}
 where $h(R)$ is the eddy diffusivity which scales with $R$ as a power law
 $R^{\zeta_h}$.  Can we use this to guess a form for $D_n(R)$ in the
 present case? On the face of it the answer is no. Our evaluation
 (\ref{eval1}) indicated that if we have a differential operator it
 operates on $S_{n+1}$ rather than on $S_n$. On the other hand, once we
 assume that the leading order evaluation cancels in $D_n$, the next order
 is again in terms of $S_n$ as in the case of passive scalar. In fact, it
 is not difficult to see (Appendix A) that the topology of the weakly
 linked diagrams for $D_n$ is identical (after the cancellation of the
 leading order) to the topology of the leading contributions for $D_n$ in
 the case of the passive scalar. We thus guess that for the aim
 of evaluation of the coefficient we can write $D_n(R)$ with an
 eddy-viscosity similar to (\ref{passD}) in which $h(R)$ is found by
 comparing (\ref{passD}) and (\ref{eval2}):
 \begin{equation}
 D_n(R) = b  {1\over R^2}{d\over dR}R^2 {S_3(R) R\over S_2(R)}
  {d\over dR}S_n(R) \ , \label{NSD}
 \end{equation}
 with $b$ being now an $n$-independent coefficient. The physical meaning of
 this guess is that the $R$-dependent eddy viscosity $h(R)$ takes here the
 form
 \begin{equation}
 h(R) = b {RS_3(R)\over S_2(R) }\ . \label{eddy}
 \end{equation}
 Note that the eddy diffusivity which is introduced here scales like
 $R^{2-\zeta_2} = R^\Delta$.
 \subsubsection{Scaling Exponents in the Eddy-Viscosity Approximation}
 Using (\ref{NSD}) we compute
 \begin{equation}
 d_n(\zeta_n) = b \zeta_n (3+\zeta_n-\zeta_2) \ , \label{dn1}
 \end{equation}
 where we used the fact that $\zeta_3=1$. Together with (\ref{coefbal}) we
 find a quadratic equation for $\zeta_n$:
 \begin{equation}
 2 b \zeta_n (3+\zeta_n-\zeta_2) = n\ . \label{dn2}
 \end{equation}
 We remind the reader that this equation is valid for $n>2$ since there is
 no cancellation of the leading scaling order in $D_2$. Using again the
 fact that $\zeta_3=1$ we find the constraint
 \begin{equation}
 2 b (4-\zeta_2)=3 \ . \label{constraint}
 \end{equation}
 This leaves us in (\ref{dn2}) with only one unknown number which we take
 as $\zeta_2$. Solving for $\zeta_n$ we find
 \begin{equation}
 \zeta_n = {3-\zeta_2\over 2} \left[ -1+\sqrt{1+{4n(4-\zeta_2)
 \over 3(3-\zeta_2)^2}}\right]
  \ . \label{model}
 \end{equation}
 Not having a theoretical value of $\zeta_2$ we can take for it any of the
 following values: (i) The K41 value $\zeta_2=2/3$. (ii) The experimentally
 ``accepted" value $\zeta_2=0.70$. (iii) We can assume that Eq.
 (\ref{dn2}) is correct also for $n=2$. This gives $\zeta_2=8/11\simeq
 0.727$. These different choices lead to the numerical values of $\zeta_n$
 shown in Table 1 under the rows  labelled ``model (i), (ii) and (iii)".

 In the same table we also show the result of three available experiments.
 It should be noted that all these experiments refer to Reynolds number
 Re$_\lambda<10^3$, and thus as discussed above, our asymptotic predictions
 may not apply. Corrections to our asymptotic behaviour are still too large
 to be ignored. This may be the reason for the deviations between the
 theoretical predictions and the experimental numbers.
 \section{Concluding remarks for papers I-III and the road ahead}
 We believe that the theory that was presented in papers I-III contains
 novel elements that are likely to remain as cornerstones in the theory of the
 fine structure of hydrodynamic turbulence. Since the approach is highly
 technical, we attempt in these concluding remarks to summarize first what
 are the essential elements of the analysis both from the point of view of
 technique and of physics.

 One fundamental issue that separates our approach from the traditional
 K41 approach is the explicit appearance of anomalous exponents with the
 ultraviolet cutoff length as the renormalization scale. It is intuitively
 clear that gradient fields cannot be insensitive to $\eta$ since by
 definition a gradient measures differences on scales smaller than $\eta$,
 and this length may appear when one attempts to calculate the correlation
 functions of gadient fields. Indeed, an important step in our theory is
 the explicit identification of the ultraviolet divergences which appear
 in the diagrammatic theory of correlations of gradient fields, and the
 computation of the largest anomalous exponent associated with such
 divergences.  With this in hand we could go further in nonperturbative
 analysis and finally discover the mechanism for anomalous scaling. It is
 not completely evident that our route is either economical or unique.
 This route reflects the nature of our understanding and the path followed
 by us, and it is possible that there is a shorter path to the same
 results.  Only future research will tell us about that.

 Let us summarize the technical aspects of our analysis.  The foundation of
 the theory is the Navier Stokes equations.  Starting from these equations
 we use a combination of renormalized  (but order by order) analysis of
 diagrammatic perturbation series with some  exact, non-perturbative
 considerations. We found that only the latter allowed us to derive
 multiscaling. Order by order analysis could not take us out of the
 traditional K41 scaling laws.

 The scale invariant formulation of diagrammatic perturbation theory for
 fluid mecahnics to all orders becomes technically  tractable due to the
 Belinicher-L'vov transformation of the Eulerian velocity field. Without it
 all known approaches are either plagued by infrared divergences or are
 limited to low order perturbation terms. In paper I we developed a Wyld
 type diagrammatic technique for BL velocity differences in ${\bf r},\,
 t$-representation. This presentation allows one to consider a $p$-th order
 diagram for the structure function $S_n(R)$  as a set of ``elementary"
 interactions involving $p$ intermediate points with integration over their
 space-time coordinates. One important discovery of paper I is the {\it
 property of locality} which means that the major contribution to these
 integrals originates from a {\it ball of locality} which is a sphere of
 radius of the order of $R$. The second important fact which has
 implications on the structure of the perturbation theory is the property
 discovered in paper II which we called {\it rigidity} of diagrams.  To
 understand this property one may use an analogy between diagrams and a
 mechanical system in which the propagators are substituted by springs
 which are joined at the vertices.  The point is that the dominant
 contribution to the diagrams for any object originates from the region of
 ``mechanical equilibrium" of the complementary mechanical sytem. In the
 mechanical analog the external end of springs (propagators) are fixed at
 the external coordinates of the object discussed. The property of rigidity
 is very helpful in analysing the asymptotic behavior of many point objects
 when some coordinates are much larger (or smaller) than others.  It turned
 out that such asymptotic considerations for many point correlation
 functions play an important role in understandng the mechanism that leads
 to anomalous exponents.

 The main result of Paper II is the demonstration of the existence and
 origin of anomalous exponents in the analytical theory of turbulence.As
 in the theory of second order phase transitions there is a logarithmic
 divergence in the UV regime of ladder diagrams for correlation functions
 of velocity {\it gradients}.Anomalous behavior appears as a
 non-perturbative solution of some formally exact diagrammatic equation
 which one obtains after resumming the ladder diagrams.  Physically,
 anomalous behavior of correlation functions with two (or more) very
 different separation distances $r\ll R$ is the result of a summation over
 a large number ${\cal N}=(R/r)^\Delta$ of equally important channels of
 interaction of  turbulent motions having scales between $r$ and $R$.  In
 paper II we succeeded to compute the anomalous scaling exponent $\Delta$
 and found that $\Delta$ exactly equals its critical value $\Delta_{\rm
 cr}=2-\zeta_2$. This fact is of crucial importance for the nature of
 anomalous scaling in hydrodynamic turbulence.  In the hypothetical case
 $\Delta<\Delta_{\rm cr}$  one  expects  K41 scaling of the structure
 functions in the limit Re$\to \infty$. There can be only subcritical
 corrections to this, and such corrections  have the  {\it viscous scale}
$\eta$ as the renormalization scale.  In reality   $\Delta=\Delta_{\rm
 cr}$  and this opens up the possibility for the destruction of K41  for
 all values of Re.

 The analysis of the possible mechanisms for anomalous (non-K41) scaling
 of the structure function is the main topic of Paper III. This analysis
 is based on the exact and nonperturbative {\it balance equations}
 $D_n(R)=J_n(R)$ which follow from the equation of motion for the
 structure functions $S_n(R)$.  These equations are a direct consequence
 of the Navier-Stokes equations in the statistical stationary state  in
 which $\partial S_n(R,t)/\partial t=0$.  We succeeded to compute (in the
 limit Re$\to \infty$) the viscous term $J_n(R)$ exactly:  $J_n(R)= n J_2
 S_n(R)/2 S_2(R)$. This result is the leading term among a few terms, and
 is the contribution of strongly linked diagrams. It is proportional to
 $R^{\zeta_n-\zeta_2}$ as a direct consequence of the criticality of the
 theory in the sense that $\Delta=2-\zeta_2$.  This contribution is much
larger than the contribution of non-linked diagrams, which are the result
 of Gaussian decompositions. The largest of these next order terms is
 proportional to $R^{\zeta_{n-2}}$. In the hypothetical subcritical case
 when $\Delta<2-\zeta_2$  the latter term dominates the strongly linked
 terms and classical K41 scaling is then unavoidable.

 Because of the importance of this point we reiterate: because $\Delta$
 attains its critical value $2-\zeta_2$ the dominant contribution to the
 viscous term $J_n(R)$ is proportional to $S_n(R)/S_2(R)$ and not to
 $S_{n-2}(R)$. This is one of three findings which open the way to
 multiscaling in the theory of hydrodynamical turbulence. The second one
 is less solid: it is the {\it assumption} that the leading (stronly
 linked) contributions to the interaction term $D_n(R)$ cancel exactly. We
 have only some arguments as to why this may happen. The detailed analysis
 of such a possibility and the clarification of the possible relation of
 the cancellation with (hidden) symmetries of fluid mechanics are
 important aspects of future research.  What we propose now is that this
 cancellation is {\it the only visible} mechanism to induce multiscaling
 in the theory of the structure functions. It is very possible that this
 is the only mechanism for multiscaling, and if we accept the opinion of
 the majority of the workers in the field of turbulence that experiments
 indicate the existence of multiscaling, we are forced to accept that
 there is a cancellation of the strongly linked contribution to the
 interaction term $D_n(R)$.  Lastly, the third finding leading to
 multiscaling is the understanding of the important role of weakly linked
 contributions to $S_n(R)$ which scale like
 $R^{\zeta_{n-1}+\zeta_3-\zeta_2}$. In a multiscaling situation this term
 dominates over the leading irreducible contribution which is proportional
 to $R^{\zeta_{n-2}+\zeta_2}$.  If one misses the weakly linked terms, K41
 scaling reappears again.

 Besides issues of principle, there remains the actual calculation of the
 scaling exponents $\zeta_n$. To accomplish such a calculation we need to
 know the {\it coefficient} in the evaluation of $D_n$. From the ``eddy
 viscosity" approximation for $D_n$ we were led to the exponents shown in
 Table 1. We cannot say that the agreement with the experimental data is
 amazing. The discrepancies may stem from various sources.  Firstly, it is
 possible that the ``eddy viscosity" approximation misses something
 important in the $n$ dependence of the coefficient of $D_n$. This is
 possible, but may be not the main reason for the discrepancy. After all,
 in our theoretical developments we used very strongly the limit Re$\to
 \infty$. This limit was assumed for example in neglecting the
 disconnected contributions to $D_n$ in favor of the weakly linked
 contributions. The difference in the $R$ dependence between these two
 types of terms (in the case of $D_4$ for example) is of the order of
 $R^{0.15}$. We thus need at least 3-4 decades of inertial range to
 justify this procedure. It is very possible that the experimental results
 are still suffering from the effects of subleading contributions, and
 only future analysis that takes such contributions into account may shed
 further light on the issue.

 One particularly pressing subject for future research is the asymptotic
 behaviour of $n$-order correlation functions. We found in the present
 paper that when two of the coordinates of $F_n({\bf r}_1,{\bf
 r}_2,\dots{\bf r}_n)$ (say ${\bf r}_1$ and ${\bf r}_2$) coalesce, the
 correlation function is proportional to $S_2(r_1)$. This is just one
 example of the asymptotic properties that are summarized under the term
 ``fusion rules". We need to understand the deep structure of the theory
 that is responsible for this fusion rule, but that will also allow us to
 predict what happens, say, when three or more points coalesce. We guess
 that the $n$-point correlation function in that case will be proportional
 to $F_3$, etc. Formally we need to develop the operator algebra that will
 automatically furnish all the needed fusion rules, and will be also
 compatible with multiscaling. We plan to propose elements of such a theory
 in a forthcoming publication \cite{95LP-e}.

 Finally, we comment on the relation and differences between scaling in
 turbulence and scaling in better understood subjects like critical
 phenomena. Because of the superficial similarities (many body problems
 with strong interaction and scale invariance) there were many attepmts to
 apply formal schemes in the wake of critical phenomena to understand
 turbulence: renormalization groups, $\epsilon$-expansion, $1/d$-expansion,
 $1/N$-expansions, and what not. If the approach taken in this series of
 papers turns out to be correct, this will mean that the theory of
 turbulence is significantly different from critical phenomena.  Some
 elements reappear: sums of ladder diagrams contribute anomalous exponents,
 fusion rules are needed, and the smell of operator algebra is there.

 However that are at least two major differences: there exists flux
 equilibrium instead of thermodynamic equilibrium, and the interaction in
 the theory of turbulence is highly non-local because of the effects of
 pressure. In contrast, in critical phenomena it is sufficient to have
 local interactions that build up to global criticality because of the
 cancellation of energetic and entropic contributions to the free energy.
 In turbulence, notwithstanding the nonlocality of the interaction it
 turns out that the diagrammatic theory in BL variables is finite order by
 order. In contrast, the perturbative analysis of critical phenomena leads
 to divergences that result (after renormalization) in anomalous scaling.
 Thus the mechanism for anomalous scaling in turbulence must be different.

 Due to the flux equilibrium there is a global connection between the
 largest and smallest scales in the problem. A deep consequence of the flux
 equilibrium is the 2-point fusion rule that was discussed above as one of
 the cornerstones for multiscaling. In addition, flux equilibrium and the
 need to satisfy boundary conditions at the two ends of the energy cascade
 introduces the possibility of having the outer scale as the
 renormalization length {\it without} infrared divergences in order by
 order expansions for the structure functions.

 As a result of all these difference we do not have the fixed point
 structure with a small number of unstable directions that is so typical of
 critical phenomena. In some sense, the independence of the perturbative
 terms from a typical scale means that we have infinite number of {\it
 marginal} operators. The resummation of the perturbative theory results in
 a possibility of dressing these marginal operators, and there can be
 infinitely many independent exponents. Whether such a concept can be
 turned into a computational scheme is a question for the future.

 \section{Appendix: Weakly Linked Contributions to the interaction term}

 In this appendix we discuss the weakly linked contributions to the
 interaction term $D_n$.  In Eqs. (\ref{D2mfinal}) and (\ref{D2vecfinal})
 the integrals depend on the $n+1$-order correlation function $T_{n+1}({\bf
 r},{\bf r},{\bf R})$ in which the two first coordinates ${\bf r}$ are
 special (they are either ${\bf r} _1$ or ${\bf R} +{\bf r}_1$ and there is
 a derivative with respect to ${\bf r}_1$). In its turn every weakly linked
 contribution to $F_{n+1}({\bf r}_1,{\bf r}_2,\dots{\bf r}_{n+1})$ (or to
 $T_{n+1}({\bf r},{\bf r},{\bf R}) = F_{n+1}({\bf r},{\bf r},{\bf R},{\bf
 R},\dots {\bf R})$ has two weakly linked legs (denoted in Fig.10a as
 $x_a,x_b$), connected to the body of the diagram via a one-propagator
 bridge. There are $C^2_{n+1} = n(n+1)/2$ weakly linked contributions to
 $F_{n+1}$ in which the role of weakly linked legs is played by each pair
 taken from the $n+1$ total number of legs. The $C^2_{n+1}$ contributions
 to $D_n$ can be subdivided into three groups. The first group consists of
 just one term in which two special coordinates in $T_{n+1}$ are exactly
 the coordinates of the two weakly linked legs (${\bf r}={\bf r}_a={\bf
 r}_b$). The second group of terms in $D_n$ has $2(n-1)$ terms in which
 just one of the special coordinates in $T_{n+1}$ (but only one of the two)
 is associated with a weakly linked leg. The second special coordinate is
 free to be associated with any of the $(n-1)$ legs of the body of the weakly
 linked diagram for $F_{n+1}$. This body is an $n$-point object (see
 Fig.10a) in which one leg is used to create the bridge. The last (third)
 group of terms in $D_n$ has $C_{n-1}^2=(n-1)(n-2)/2$ terms in which two
 special $T$-coordinates may be chosen from the coordinates of any $(n-1)$
 free legs in the body of the weakly linked diagram for $F_{n+1}$.

 The first two groups of terms correspond exactly to the topology of the
 digrammatic representation of the interaction term $D_n$ in the problem of
 turbulent advection of a passive scalar field $T({\bf r},t)$, and cf.
 section VB in \cite{94LPF} and section IIB2 in \cite{95FGLP}. Consider
 Fig.10 of \cite{94LPF}. The dashed lines in this figure represent 2-point
 velocity correlators, and these are replaced in our case by wavy
 correlator lines. The wavy lines in the passive scalar case represent two
 point scalar correlators, and they remain as wavy lines in the present
 case.  The fragment of the diagram in this figure which is placed to the
 right of the legs denoted by $q_2,\ q_3,\ q_4$ and to the right of the
 vertex between $k$ and $q_1$ must be interpreted now as a contribution to
 the strongly-linked four-point velocity correlator. The last serves a a
 four- point body of a weakly linked contribution to a $5$-point velocity
 correlator $F_5$. We thus conclude that the topology of the diagrams for
 $D_n$ in the case of turbulent passive advection and the first two groups
 of weakly linked diagrams for $D_n$ in the case of Navier-Stokes
 turbulence is the same.

 It appears that the third group of $C_{n-1}^2$ terms which we described
 above forms a major difference between the problems of turbulent advection
 and Navier Stokes turbulence. In fact this group does not contribute. In
 the case of passive scalar this group is absent because of the zero value
 of the $\left<Tv\right>$ correlator. It is remarkable that in the present
 case of Navier-Stokes turbulence this group cancels under the same
 condition of cancellation of the leading (strongly linked) contributions
 to $D_n$. These terms may be considered (after severing the bridge to the
 weakly linked fragment) as strongly linked contributions to $D_{n-1}$.
 They must cancel if the scenario leading to multiscaling is assumed.

 The conclusion of this appendix is far from being trivial, and in some
 sense is very surprising.  It says that although the passive advection
 problem is linear and local, whereas Navier Stokes in nonlinear and
 non-local (pressure!), it appears that if multiscaling is expected, the
 topology of the diagrams for $D_n$ is very similar in the two cases. If
 this is correct, it must be related to some deep symmetry. If so, the Eddy
 Viscosity Approximation used in section 7 may contain some essential
 aspects of the truth.

 \widetext
 \begin{table}
 \begin{tabular}{l|c|c|c|c|c|c|c|c|c|c|c|c|c|c|c|}\\ \hline
 source&n=2&4   &5   &6   &7   &8   &9   &10  &11  &12  &13
 &14  &15  &16  &17 \\ \hline
 expt. \cite{94BCBC}&0.70&1.28&1.54&1.78&2.00&2.23
 &--&--&--&--&--&--&--&--&
 --\\ \hline
 expt. \cite{93SSJ}&0.70&1.20&1.52&1.62&1.96&2.00&2.36
 &2.36&2.70&2.68&3.08&3.02&3.48&--&
 --\\ \hline
 expt. \cite{95HW}&0.71&1.30&1.57&1.82&2.06&2.25&2.46
 &2.60&2.80&2.92&3.08&3.19&3.35&3.45&
 3.58\\ \hline
 model (i) &.667 &1.243&1.463&1.667&1.856&2.035&2.204
 &2.365&2.519&2.667&2.809&2.946
 &3.079&3.208&3.333\\ \hline
 model (ii) &.700 &1.242&1.462&1.666&1.854&2.032&2.200
 &2.360&2.514&2.661&2.803&2.939&
 3.072&3.200&3.320\\ \hline
 model (iii) &.727 &1.242&1.461&1.663&1.851&2.029&2.197
 &2.357&2.509&2.656&2.797&2.933&
 3.065&3.193&3.317\\ \hline
 \end{tabular}
 \end{table}
 \narrowtext
 \newpage~\newpage
  \begin{figure}
  \epsfxsize=8.6truecm
  \vspace{.4cm}
  \caption{Diagrammatic representation of the 4-point Green's function
  ${\cal G}_{2,2}$ defined by Eq. (3.3). Panel (a): ${\cal G}_{2,2}$ as a
  sum of weakly linked contributions ${\cal G}_{2,2}^{\rm wl}$ (shown in
  Panels (b--d)) and strongly linked contribution ${\cal G}_{2,2}^{\rm sl}$
  (shown on Fig. 2). Panel (b):${\cal G}_{2,2}^{\rm wl}$ is presented in
  terms of the three-point Green's function ${\cal G}_{2,1}$, defined by Eq.
  (3.1), and the dressed vertex A. Panels (c,d) give the diagrammatic
  representation of ${\cal G}_{2,1}$ in terms of the propagators $G$ and
  $F_2$ and the dressed vertices A and B. For the diagrammatic expansion of
  the dressed vertices A, B, and C -- see Fig. 7 of paper II. Note the new
  notation of an empty little circle in the object $\cal D$. This circle
 designates a vertex that can be either bare or dressed.}
 \label{95LP-01}
  \end{figure}
  \begin{figure}
  \epsfxsize=8.6truecm
  \vspace{.4cm}
  \caption{The diagrammatic presentation of the strongly linked
 contribution ${\cal G}_{2,2}^{\rm sl}$ to the four-point Green's function
 ${\cal G}_{2,2}$, (see Fig.1). Diagrams 1 and 2 are the Gaussian
 decomposition of ${\cal G}_{2,2}^{\rm sl}$, diagrams 3 and 5 are ladders
    with one rung (of two types), and diagrams 3 and 6 are the ladders with
    two and three rungs respectively. In contrast to the expansion in Fig.9
 of paper II for $G_2$, here one has rungs of three types. These are
 $\Sigma_{3,1}$ like in diagram 5, $\Sigma_{2,2}$ as in diagrams 3 and 4,
 and rung $\Sigma_{1,3}$ as the first rung in diagram 6. The diagrammatic
 expansion of $\Sigma_{2,2}$ is shown in Fig.3, and the expansion of
 $\Sigma_{3,1}$ and $\Sigma_{1,3}$ in Fig.4. The exact resummation of this
 series is shown in Fig. 5.}
  \label{95LP-02}
  \end{figure}
  \begin{figure}
  \epsfxsize=8.6truecm
  \vspace{.4cm}
  \caption{Diagrammatic expansion for the mass operator $\Sigma_{2,2}$
  which is found in the expansion of ${\cal G}_{2,2}^{\rm sl}$ shown in
 Fig. 2. This mass operator is an important element in the system of
 equations for ${\cal G}_{2,2}^{\rm sl}$ and ${\cal G}_{3,1}^{\rm sl}$
 shown in Fig. 5.  The principal cross sections are shown as dashed lines.}
 \label{95LP-03}
  \end{figure}
  \begin{figure}
  \epsfxsize=8.6truecm
  \vspace{.4cm}
  \caption{Diagrammatic expansions of the mass operator $\Sigma_{3,1}$
  (panel a) and $\Sigma_{1,3}$ (panel b) which appear in the expansion of
 ${\cal G}_{2,2}^{\rm sl}$ shown in Fig. 2.  These mass operators are the
 essential elements in the equations shown in Fig. 5. }
 \label{95LP-04}
  \end{figure}
  \begin{figure}
  \epsfxsize=8.6truecm
  \vspace{.4cm}
  \caption{The exact system of equations for the strongly linked
  contributions to ${\cal G}_{2,2}$ (panel a) and ${\cal G}_{3,1}$ (panel
 b) in terms of the propagators $G$ and $F_2$ and the mass operators
 $\Sigma_{3,1}$, $\Sigma_{2,2}$ and $\Sigma_{1,3}$. The first terms in the
 expansion of $\Sigma_{m,n}$ are shown in Figs.3, 4. This system results
 from the exact ressumation of ladder diagrams.}
    \label{95LP-05}
  \end{figure}
  \begin{figure}
  \epsfxsize=8.6truecm
  \vspace{.4cm}
  \caption{Diagrammatic representation of the weakly linked contributions
  to the four-point correlation function $F_4$. Panel a: a weakly linked
 contribution to $F_4({\bf r}_0|x_1,x_2,x_3,x_4)$ in which a one-propagator
 bridge is placed between (1,2) pair and the (3,4) pairs of legs. One has
 two more similar weakly-linked contributions with the bridge connecting
 the pairs (1,3) with (2,4) and (1,4) with (2,3). On the left of the
 diagrams 1 and 2 one find the objects $G_{2,1}$ and $F_3$ resepectively.
 These objects are presented diagrammatically in Fig. 1c and panel b of the
 present figure, respectively. On the right side of diagrams 1 and 2 one
 has the three-point objects that were designated as $\cal A$ and $\cal D$.
 These objects can be found in Fig.1d and in panel b of the present figure.
 }
 \label{95LP-06}
  \end{figure}
  \begin{figure}
  \epsfxsize=8.6truecm
  \vspace{.4cm}
  \caption{Panel a: diagrammatic representation of the irreducible
  four-point correlation function as a sum of weakly and strongly linked
 parts. Diagram 1 is the weakly linked part $F_4^{\rm wl}$ and diagram 2 is
 the strongly linked part $F_4^{\rm sl}$.  $F_4^{\rm wl}$ is presented in
 Fig. 6. Panel b: Exact presentation of $F_4^{\rm sl}$. The elements
 appearing in this presentation are the four-point Green's functions ${\cal
 G}_{2,2}$ and ${\cal G}_{3,1}$ on the left side and on the right side of
 the diagrams, and three different new four-point objects. The notation
 used to distinguish the three new objects is with zero, one or two
 horizontal lines inside. The diagrammatic representation of the new
 objects is shown in Fig. 8.}
     \label{95LP-07}
  \end{figure}
  \begin{figure}
  \epsfxsize=8.6truecm
  \vspace{.4cm}
  \caption{Diagramamtic representation for the three different central
  blocks which appeared in Fig. 7. These  are infinite order expansions in
 terms of contributions with increasing numbers of correlators in their
 principal cross section (i.e. 2, 3, 4  and more. The four- point obejects
 in diagram 1 in each panel are the mass operators $\Sigma_{2,2}$ and
 $\Sigma_{3,1}$ which are the dressed rungs of the ladders whose
 presentation is shown in Figs. 3 and 4.}
  \label{95LP-08}
  \end{figure}
  \begin{figure}
  \epsfxsize=8.6truecm
  \vspace{.4cm}
  \caption{An example of a fragment contribution to the right hand part of
  diagram 1 in Fig.7.}
 \label{95LP-09}
  \end{figure}
  \begin{figure}
 \epsfxsize=8.6truecm
  \vspace{.4cm}
  \caption{Diagrammatic representation of $n$-point correlation functions.
  Panel a: weakly linked contributions. Diagrams 1 and 2 are the
 generalization of diagrams 1 and 2 for $F_4$ shown in Fig.6 panel a.
 Diagrams 3 is an example of a weakly linked contribution with two
 one-propagator bridges. Such contributions do not appear in the case of
 $F_4$. Panel b:  Strongly linked contributions to $F_n$, considering the
 legs designated as 1 and 2 special.  In diagram 1 one has $F_4$ on the
 left; in diagrams 2 and 3 one has $G_{3,1}$ and $G_{2,2}$, the same
 objects that appeared in panel b of Fig. 7 for $F_4$.}
 \label{95LP-10}
  \end{figure}
  \begin{figure}
  \epsfxsize=8.6truecm
  \vspace{.4cm}
  \caption{Diagrammatic representation of the weakly linked contribution
  for $F_n$ shown as diagram 2 in Fig. 10, panel a. For the 3-point
 correlator $F_3({\bf r}_0|x_a,x_b,x_c,x_d)$ we used the representation of
 Fig. 6, panel c and placed it here in the dashed ellipse which is denoted
 ``left". On the right we have an $m$-point object ${\cal D}_m$ which is a
 generalization for the case $m>3$ of ${\cal D}_3$ which appeared in Fig.1
 panel d.}
\label{95LP-11}
 \end{figure}


\begin{thebibliography}{10}
 \bibitem[\ast]{lvov}
 L'vov's e-mail: fnlvov@wis.weizmann.ac.il.

 \bibitem[\dag]{procaccia}
 Procaccia's e-mail: cfprocac@weizmann.weizmann.ac.il.

 \bibitem{95LP-a}
 V.~S. L'vov and I.~Procaccia.
 \newblock ``Exact resummations in the theory of hydrodynamic turbulence. 0.
   {L}ine-resummed diagrammatic perturbation approach".
 \newblock In F.~David and P.~Ginsparg, editors, {\em Lecture Notes of the Les
   Houches 1994 Summer School}, 1995.
 \newblock In press.


 \bibitem{95LP-b}
 V.~S. L'vov and I.~Procaccia.
 \newblock ``Exact resummations in the theory of hydrodynamic turbulence. {I}.
   {T}he ball of locality and normal scaling".
 \newblock {\em Phys.~Rev.~E}, 1995.
 \newblock in press.

 \bibitem{95LP-c}
 V.~S. L'vov and I.~Procaccia.
 \newblock ``Exact resummations in the theory of hydrodynamic turbulence. {II}.
   A ladder to anomalous scaling".
 \newblock {\em Phys.~Rev.~E}, 1995.
 \newblock in press.

 \bibitem{41Kol}
 A.N. Kolmogorov, Dokl. Akad. Nauk SSSR, {\bf 30}, 229 (1941)

 \bibitem{MY-2}
 A.~S. Monin and A.~M. Yaglom.
 \newblock {\em Statistical Fluid Mechanics}, volume~2.
 \newblock MIT Press, Cambridge, 1975.


 \bibitem{95Nel}
 M.~Nelkin.
 \newblock ``Universality and scaling in fully developed turbulence".
 \newblock {\em Advan. in Phys.}, {\bf 43},143 (1994).

 \bibitem{Fri}
 Uriel Frisch.
 \newblock ``{\em Turbulence: The Legacy of A.N. Kolmogorov}".
 \newblock Cambridge University Press, Cambridge, 1995.
 \newblock In press.


 \bibitem{93SK}
 K.R. Sreenivasan and P.~Kailasnath.
 \newblock ``An update on the intermittency exponent in turbulence".
 \newblock {\em Phys.~Fluids~A}, 5(2):512--514, 1993.

 \bibitem{61Wyl}
 H.~W. Wyld.
 \newblock ``Formulation of the theory of turbulence in an incompressible
fluid".
 \newblock {\em Ann.~Phys.}, 14:143--165, 1961.

 \bibitem{73MSR}
 P.~C. Martin, E.~D. Siggia, and H.~A. Rose.
 \newblock ``Statistical dynamics of classical systems".
 \newblock {\em Phys.~Rev.~A}, 8(1):423--437, July 1973.

 \bibitem{78DP}
 C.~De Dominicis and L.~Peliti.
 \newblock {\em Phys.~Rev.~B}, 18:353, 1978.

 \bibitem{87BL}
 V.~I. Belinicher and V.~S. L'vov.
 \newblock ``A scale-invariant theory of fully developed hydrodynamic
turbulence".
 \newblock {\em Sov.~Phys.~JETP}, 66:303--313, 1987.

 \bibitem{95LL}
 V.~S. L'vov and V.~V. Lebedev.
 \newblock ``Anomalous scaling in {K}olmogorov 1941 turbulence".
 \newblock {\em Europhys. Lett.}, {\bf 29},681 (1995).
 \newblock in press.

 \bibitem{94LL}
 V.~V. Lebedev and V.~S. L'vov.
 \newblock ``Scaling of correlation functions of velocity gradients in
   hydrodynamic turbulence".
 \newblock {\em JETP Letters}, 59:577--583, 1994.

 \bibitem{95LP}
 V.~S. L'vov and I.~Procaccia.
 \newblock ````{I}ntermittency'' in turbulence as intermediate asymptotics to
   {K}olmogorov'41 scaling".
 \newblock {\em Phys.~Rev.~Lett.}, {\bf 74} 2690--2693 (1995).

 \bibitem{95LP-e}
 V.S. L'vov and I. Procaccia, ``Fusion Rules and Operator Algebra in
 Multiscaling Hydrodynamic Turbulence", in preparation.

 \bibitem{94Kra}
 R.H. Kraichnan, ``Anomalous Scaling of a Randomly Advected Passive Scalar",
 Phys.Rev. Lett., {\bf 72} 1016-1019 (1994).

 \bibitem{95FGLP}
 A.L. Fairhall, O. Gat, V.S. L'vov and I. Procaccia ``Anomalous Scaling in a
 Model of Passive Scalar Advection: Exact Results", Phys. Rev. E,
 submitted.

 \bibitem{78FSN}
 Uriel Frisch, Pierre-Louis Sulem, and Mark Nelkin.
 \newblock ``A simple dynamical model of intermittent fully developed
 turbulence".
  \newblock {\em J.~Fluid~Mech.}, 87:719--736, 1978.

 \bibitem{94Got}
 T. Gotoh, Phys. Fluids {bf 6}, 3985 (1994).

\bibitem{95Pol}
A.N. Polyakov, ``Turbulence without pressure", unpublished.

 \bibitem{94LPF}
 V.~S. L'vov, I.~Procaccia, and A.~Fairhall.
 \newblock ``Anomalous scaling in fluid dynamics: {T}he case of passive
scalar".
 \newblock {\em Phys.~Rev.~E}, {\bf 50}, 4684 (1994).

 \bibitem{94BCBC}
 R. Benzi,S. Ciliberto, C. Baudet and G. Ruiz Chavarria, Physica {\bf D 80},
385 (1995)

 \bibitem{93SSJ}
 G.Stolovitzky, K.R.Sreenivasan and A. Juneja, Phys.Rev. {\bf E48},R3217 (1994)

 \bibitem{95HW}
 J. Herweijer and W. van de Water, Phys.Rev.Lett. {\bf74}, 4651 (1995).
 \end{thebibliography}
 
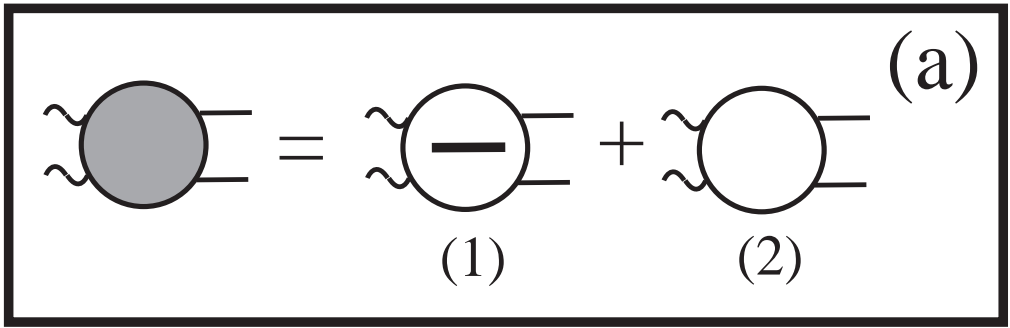
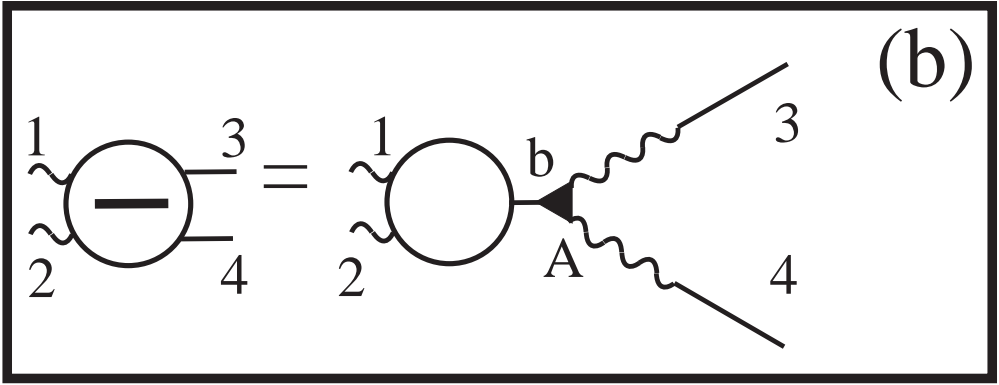
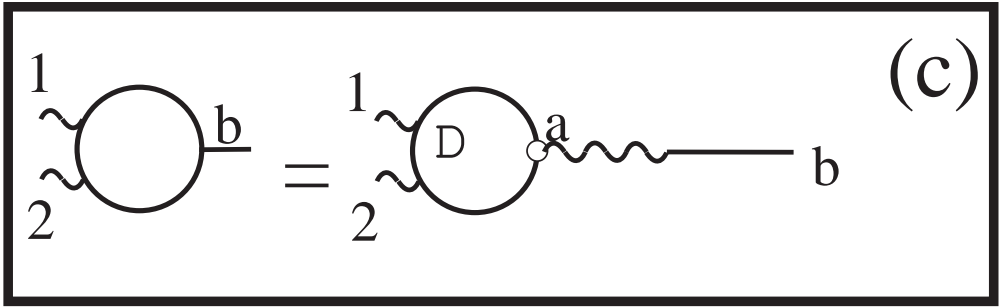
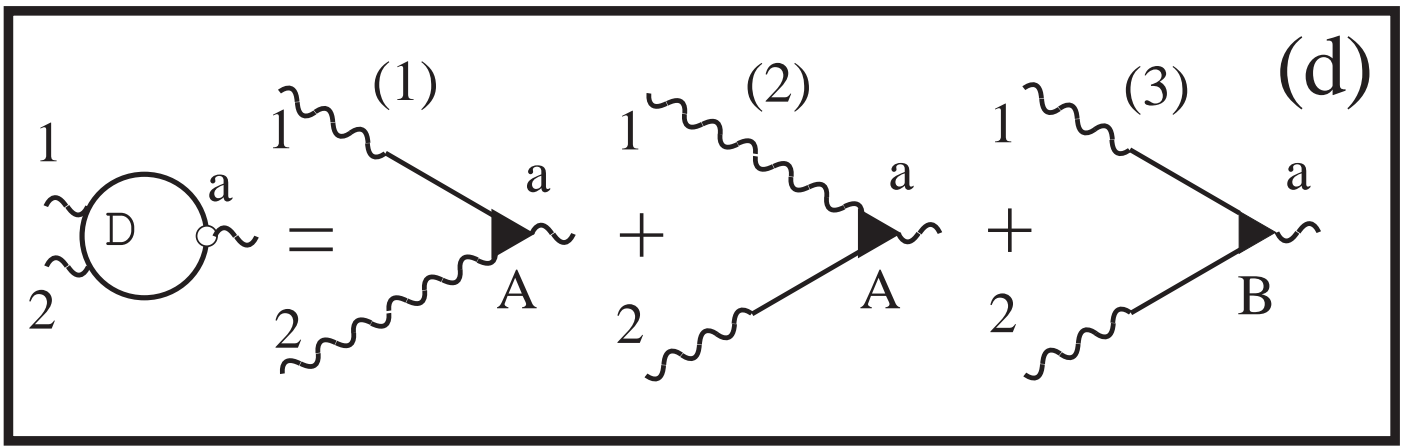
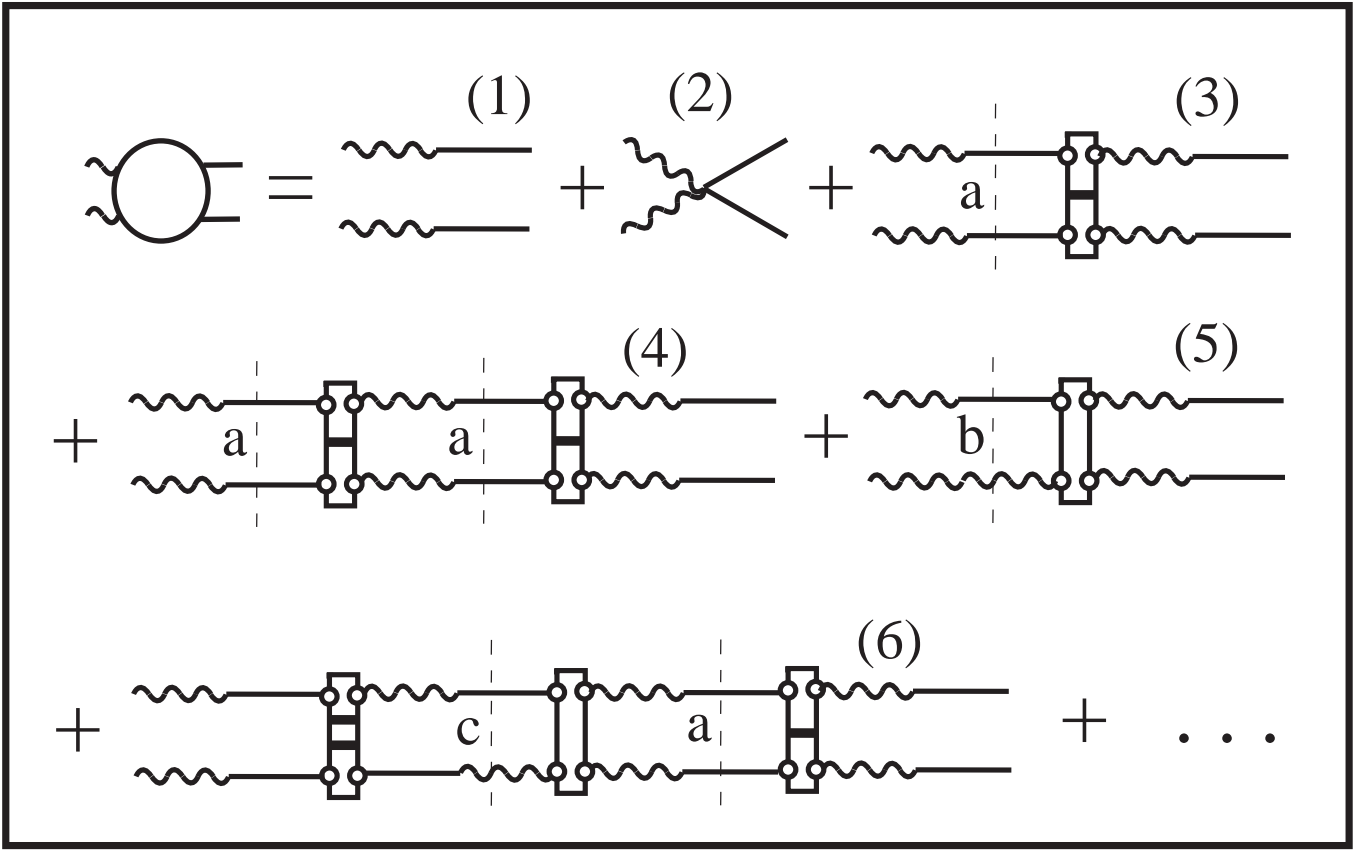
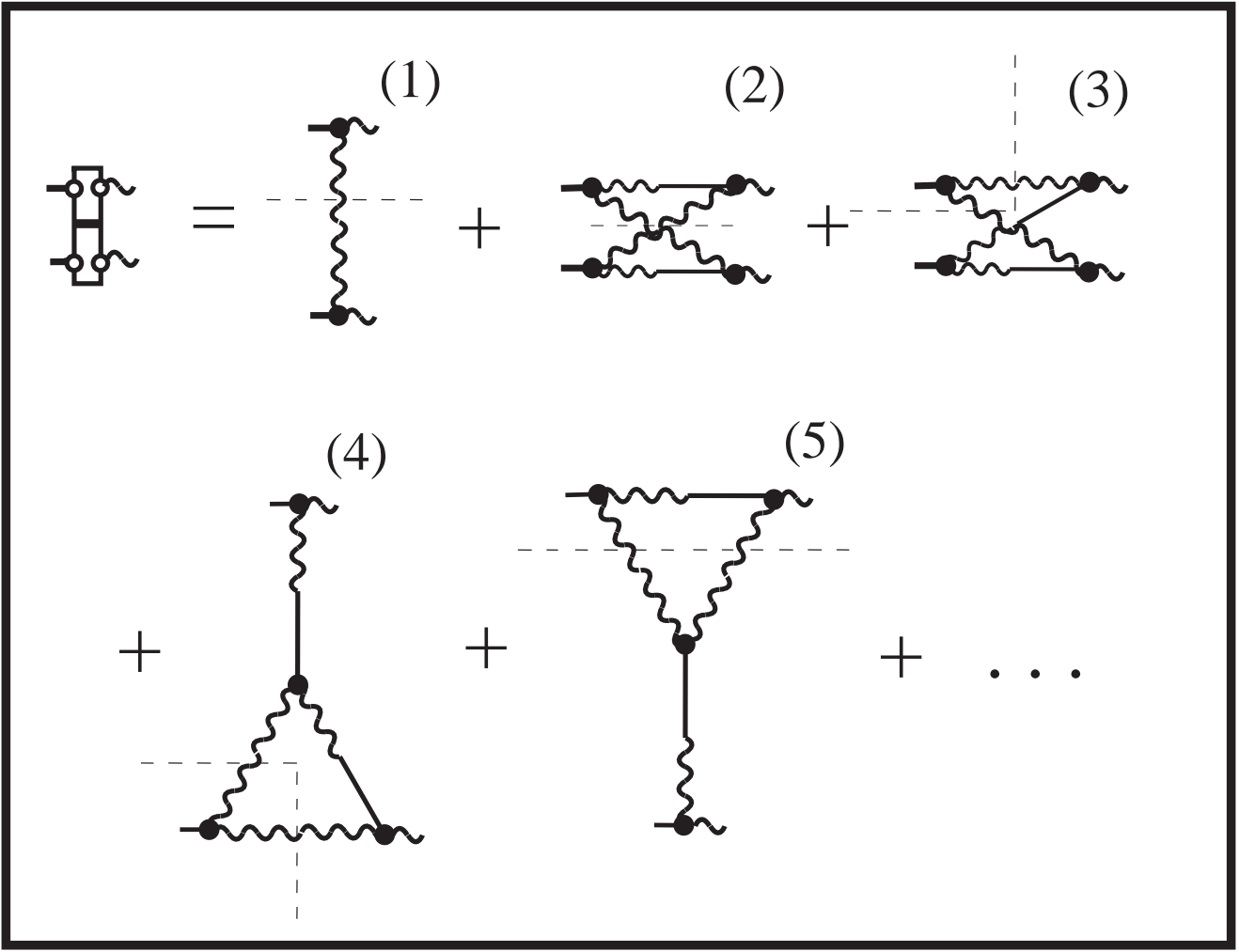
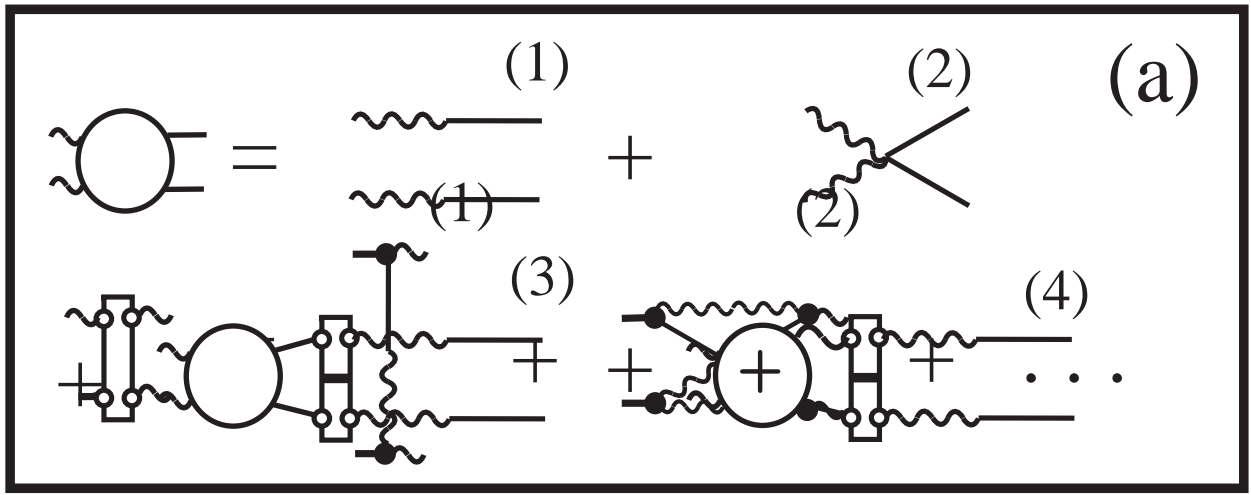
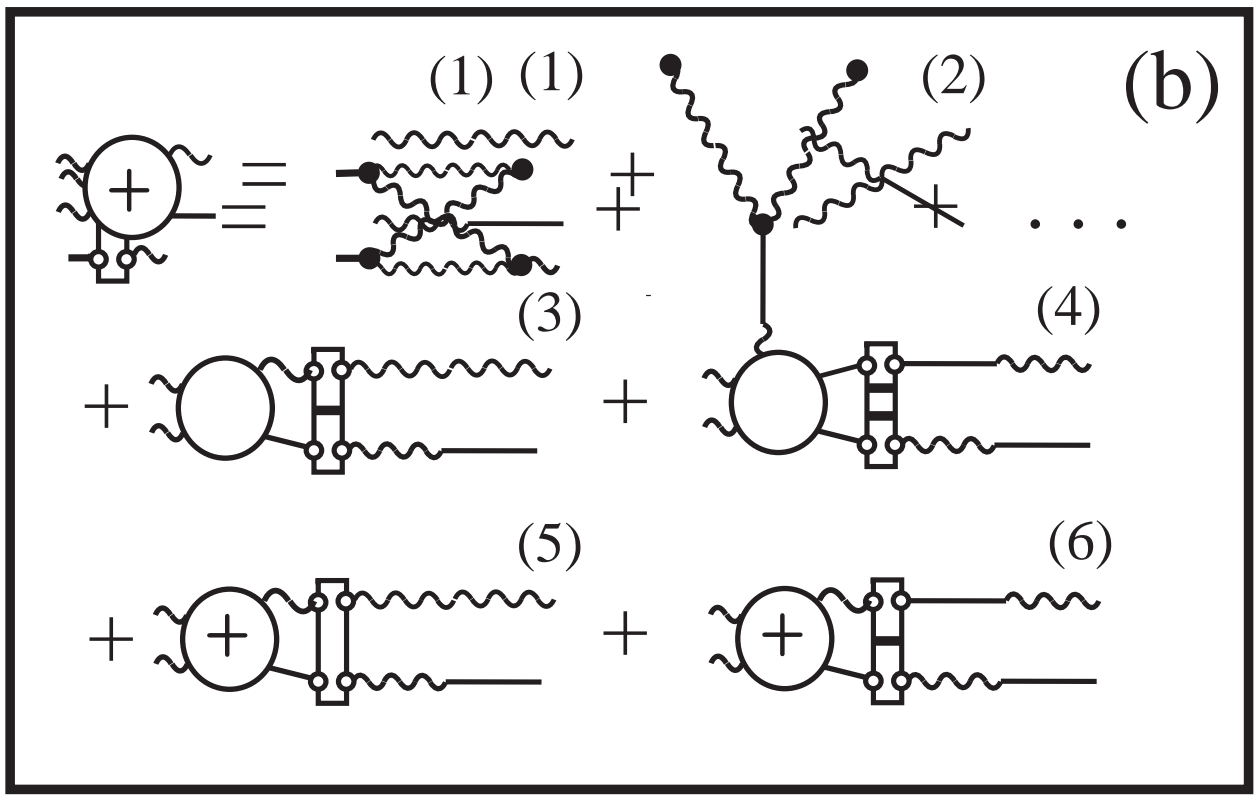
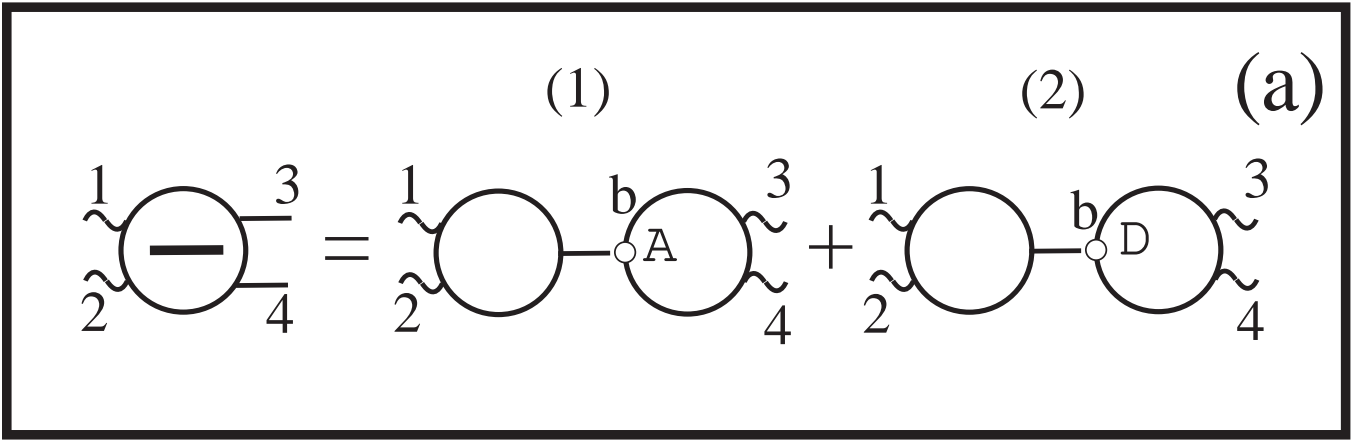
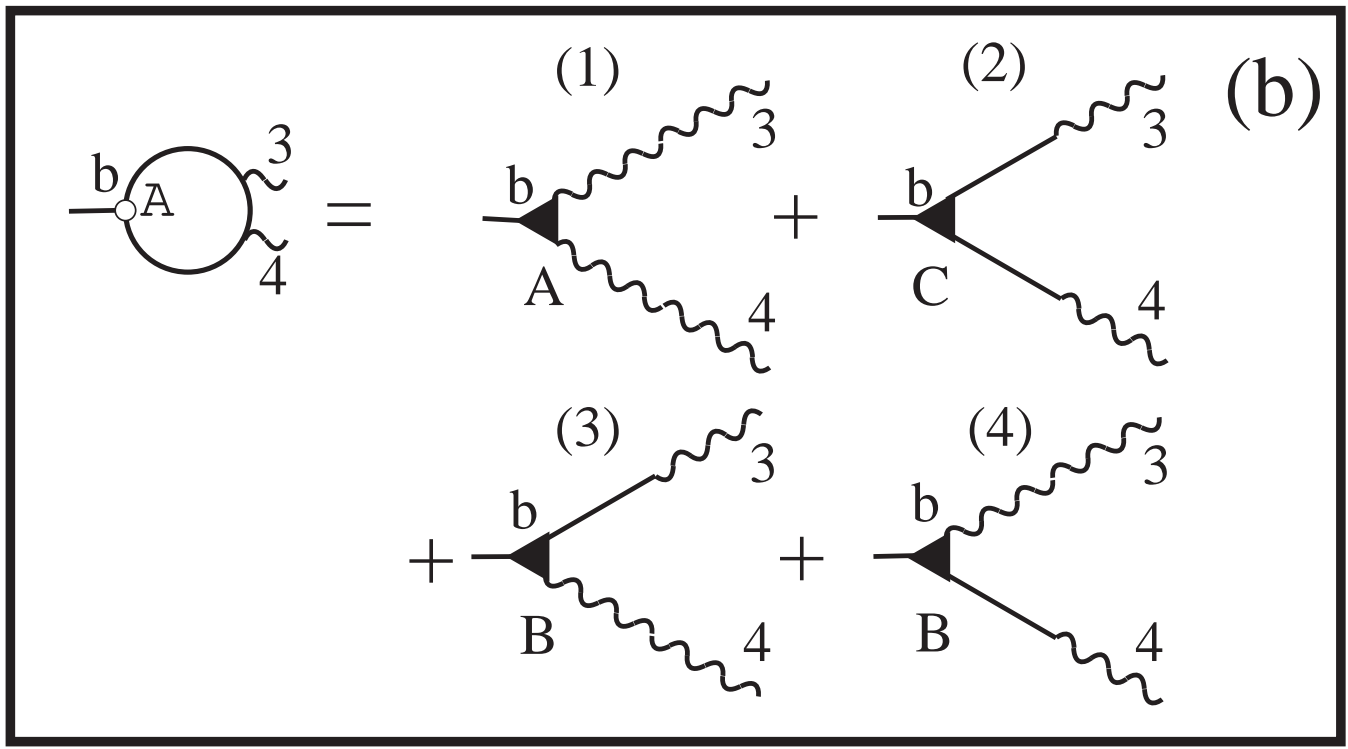
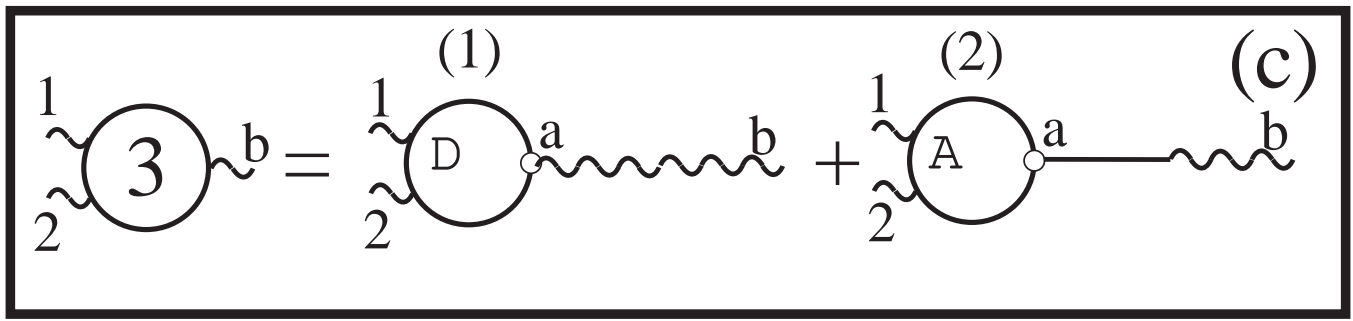
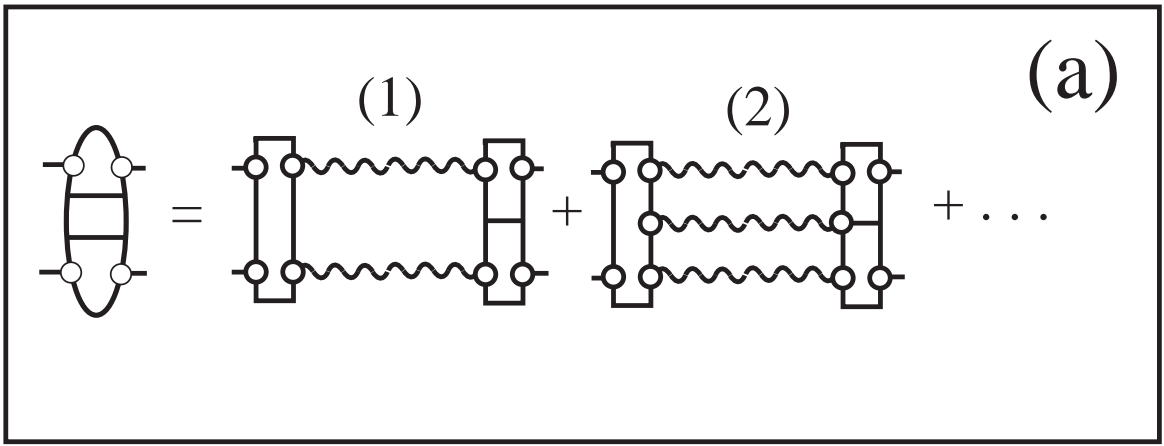
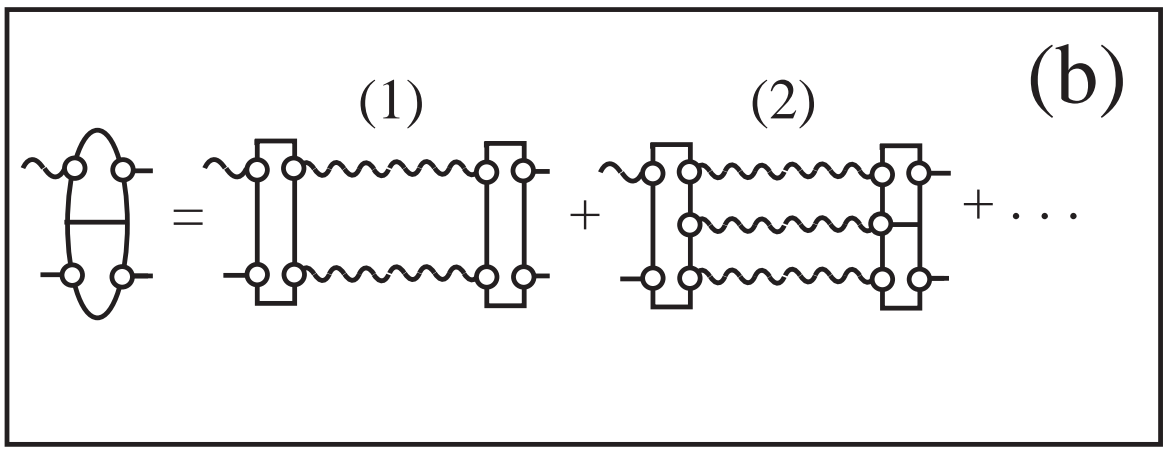
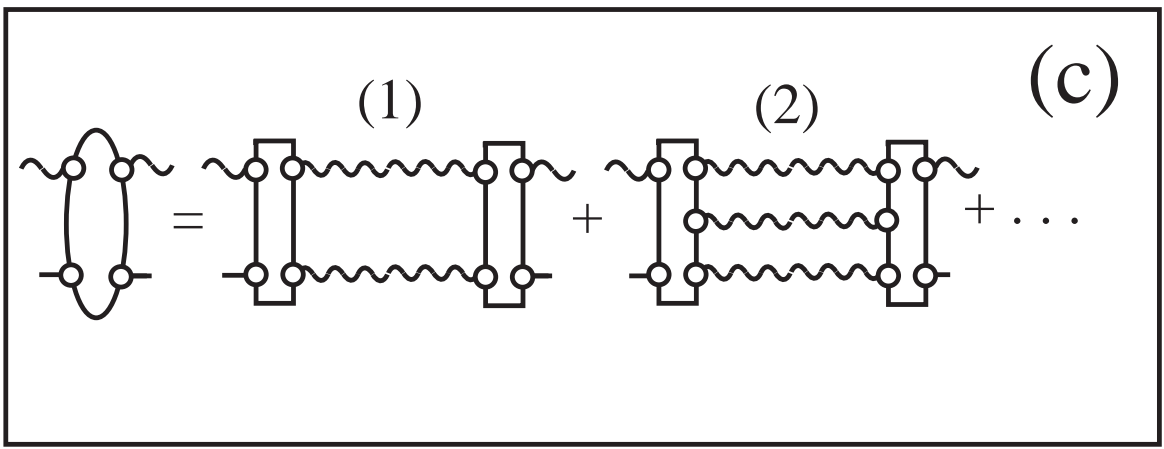
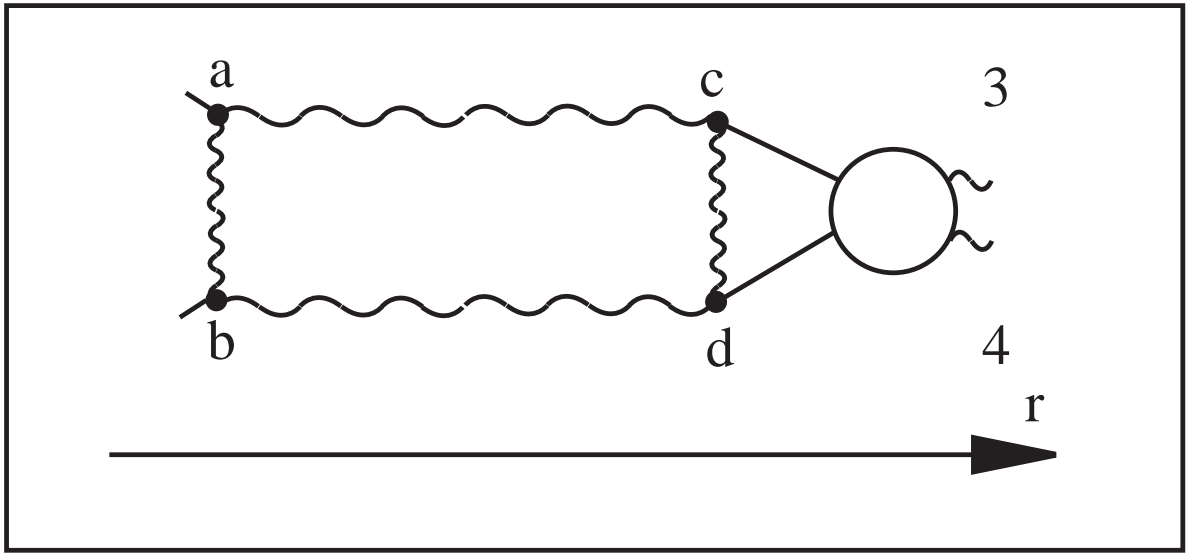
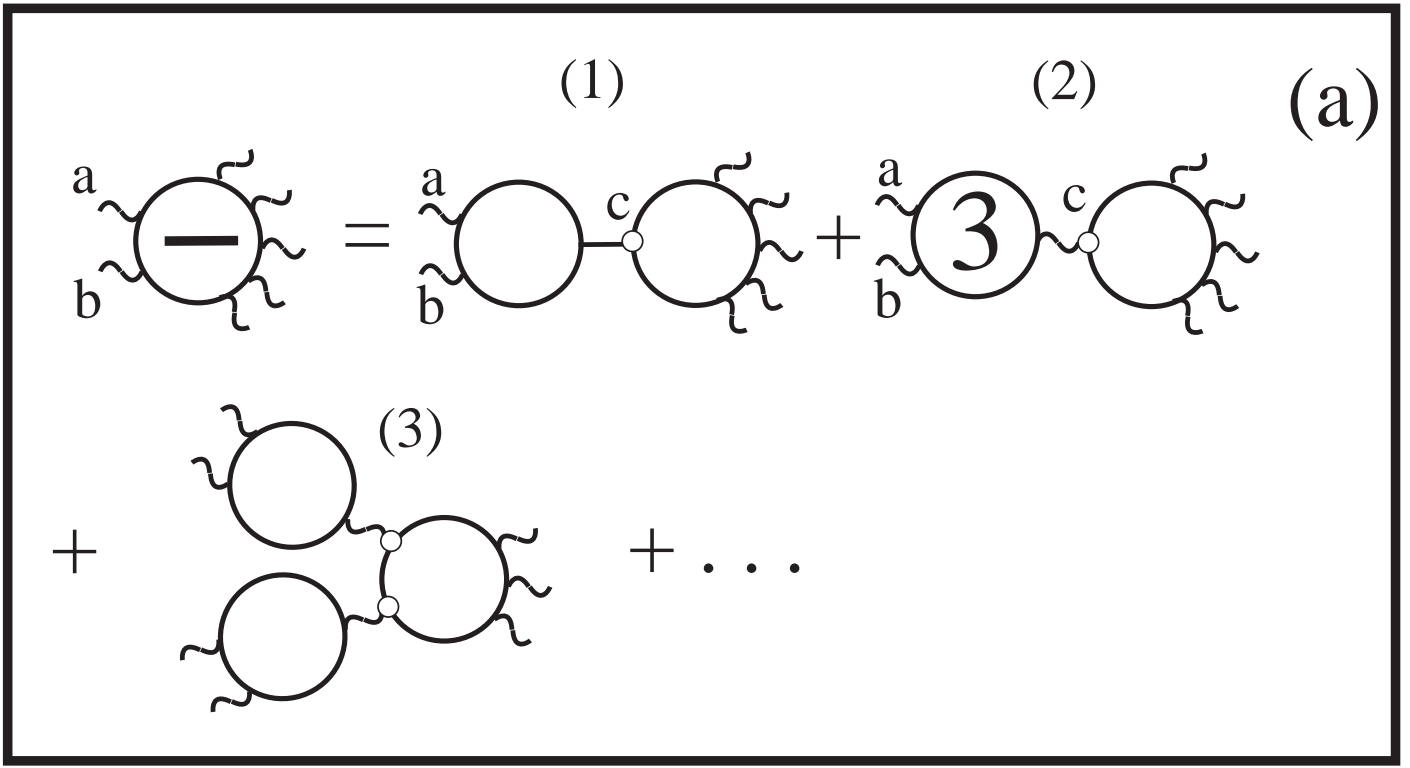
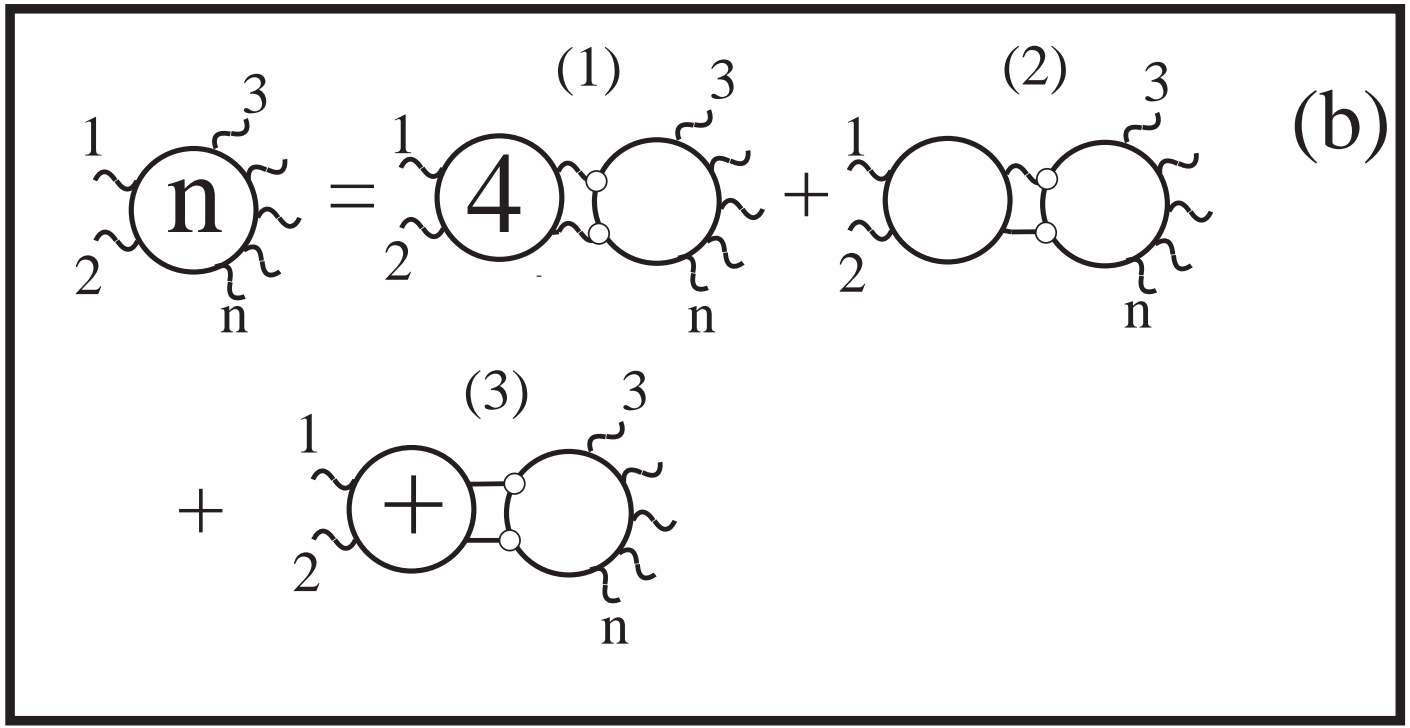
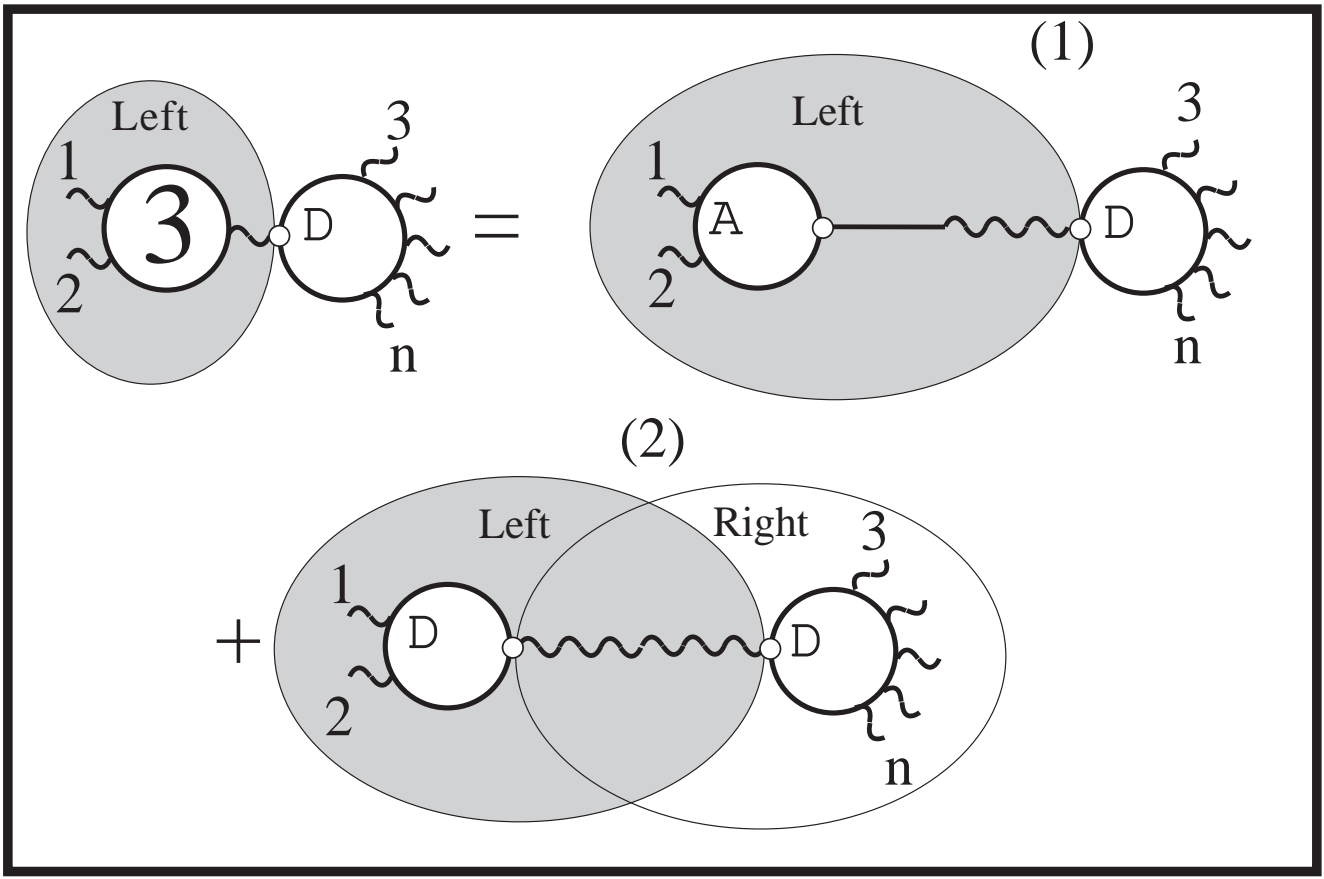
\end{document}